\documentclass[aps, pra, preprint, groupedaddress, amsfonts,
               amsmath, amssymb, showpacs, nofootinbib]{revtex4-1}
\usepackage{microtype}
\usepackage{graphicx}
\usepackage{amssymb}
\usepackage{mathtools}
\usepackage{epstopdf}
\usepackage[utf8]{inputenc}
\usepackage[T1]{fontenc}
\usepackage[usenames,dvipsnames]{xcolor}
\usepackage{hyperref}

\usepackage{color}

\def\bm#1{\mbox{\boldmath $#1$}}

\begin{document}

\title{The Independent Atom Model - Pixel Counting Method\\ for Ion--Molecule Collisions} 

\author{Hans J\"urgen L\"udde$^1$\footnote{Email: luedde@itp.uni-frankfurt.de},
Marko Horbatsch$^2$\footnote{Email: marko@yorku.ca}, and Tom Kirchner$^2$\footnote{Email: tomk@yorku.ca}}

\address{$^1$Center for Scientific Computing, Goethe-Universit\"at, D-60438 Frankfurt, Germany}
\address{$^2$Department of Physics and Astronomy, York University, Toronto, Ontario, Canada M3J 1P3}

\begin{abstract}
The independent atom model - pixel counting method (IAM-PCM) for the description of ion--molecule collisions is reviewed.
The method was introduced (in 2016) to improve on the simple additivity rule 
according to which scattering cross sections for a molecular target can be obtained by summing up the cross sections 
of the constituent atoms. The key idea of the IAM-PCM is the inclusion of weight factors in the summation,
to be determined from a geometrical interpretation
of the resultant cross section as a combined area of overlapping atomic contributions, which is calculated via
a pixelization technique.
We argue here that the IAM-PCM can be conceptualized in a different but equivalent way by associating each
pixel in the area decomposition with a scattering event. The calculation of net and charge-state correlated capture
and ionization cross sections is explained, and results for 10 keV to 10 MeV proton impact
are discussed for a number of targets ranging
from compact ten-electron systems to large biomolecules.
A previously observed scaling behaviour of the net ionization cross sections is revisited and 
shown to be captured by a simple parametrization with remarkable accuracy.
\end{abstract}

\maketitle

\section{Introduction}
\label{sec:intro}
The high level of current activity in the field of ion--molecule collisions is
a reflection of both applied and basic research interests. 
On the applied side, the relevance of these collisions in the context of radiation medicine
and ion-beam cancer therapy in particular is an often quoted motivation
for studying them (see, e.g., Refs.~\cite{Belkic_2019, Belkic_2021} and references therein).
On the more fundamental level, it is the fascination with the
inherent complexity of interacting
few-body Coulomb systems that keeps researchers drawn to the field.
This book, of course, is testament to these ongoing activities and endavours.

Building on earlier work for ion--\textit{atom} collisions, we started a research program to deal
with molecular targets in the late 2000s. A molecular-orbital (MO) 
independent-electron-model (IEM) based approach was developed and successfully
applied to collisions involving compact target molecules such as water~\cite{Murakami12a, Kirchner13} 
and methane~\cite{Arash17}.
The approach makes use of a spectral representation of the 
molecular Hamiltonian and a single-center expansion of the MOs, 
as a consequence of which it is not applicable to more extended target systems such as biomolecules
which are of obvious interest for applications in radiation medicine.
In order to be able to address such truly complex targets, we proposed an independent-atom-model (IAM) based
approach in Ref.~\cite{hjl16}. A geometrical interpretation was put forward in which the 
electron transfer (capture) and free electron production (ionization) cross sections of a molecular
target are represented as effective areas composed of overlapping atomic contributions.
Since the effective area calculations are carried out via pixelization, we termed the
approach pixel counting method and introduced the acronym IAM-PCM~\cite{hjl18}.

In Ref.~\cite{hjl16},
net capture and net ionization cross sections were presented in comparison with IAM-AR (AR stands for additivity rule) and with experimental
data for collisions of protons with small molecules (CO, $\rm H_2O$) and with uracil ($\rm C_4H_4N_2O_2$).
In Ref.~\cite{hjl18} the theory was laid out in more detail and applications were presented for water molecule clusters,
as well as for clusters of neon and carbon with a modelling of the net cross sections with cluster size.
Reference~\cite{hjl19} describes applications to a series of biomolecules (pyrimidine, purine, THF, TMP) where comparison
with electron scattering data can be made at high collision energies. The scaling behaviour of the net cross sections
with molecule size was deduced from the atomic cross sections. Tabulated data are provided for a 
large number of molecules.  
Reference~\cite{hjl19b} reports results for methane and DNA and RNA nucleobases, where in addition to experiments
comparison was also made with continuum distorted wave - eikonal initial state (CDW-EIS) calculations.
The IAM-PCM was then applied to collisions with charged (bare) ions on the basis of atomic collision cross sections
for projectile charges $Q_P=1,2,3$ in Ref.~\cite{hjl20a}. A scaling model was developed to
treat collisions with higher projectile charges, and tested up to $Q_P=13$ by comparing with experimental results 
and CDW-EIS for uracil.
In Ref.~\cite{hjl20} net capture was investigated for multiply-charged ion collisions with biomolecules
and comparison was made with a classical-trajectory Monte-Carlo (CTMC) model~\cite{PhysRevA.99.062701} for the water molecule target.
Reference~\cite{hjl22} lays out the treatment of charge-state correlated processes, i.e., goes beyond the
calculation of net capture and ionization cross sections with the IAM-PCM.
Comparison is made with experimental results for the production of singly and multiply charged target ions (which undergo fragmentation)
in proton collisions with the `isoelectronic' sequence $\rm CH_4, NH_3, H_2O$.

Recently, the IAM-PCM approach was adopted by another group, which uses time-dependent density functional theory proton--atom (net) capture cross sections
calculated with OCTOPUS~\cite{Bernal2024}. While their originally published paper shows results that deviate from our published
results the authors found an error in their code, and their results now agree quite well with our data~\cite{Bernal2025}.

This chapter summarizes the IAM-PCM approach (in Sec.~\ref{sec:theory}) and provides a sample of results for proton impact (in Sec.~\ref{sec:results}).
We mostly draw on our previous papers~\cite{hjl19, hjl22} but go beyond the material and
analysis presented there by including new results (e.g. for HF target molecules) and 
elaborating on the idea to scale the cross sections for complex biomolecules in an efficient way.
The chapter ends with concluding remarks in Sec.~\ref{sec:conclusions}.

\section{Theoretical Considerations}\label{sec:theory}
We discuss ion--molecule collisions in the semiclassical
impact-parameter framework: The projectile is assumed to
travel on a straight-line classical path ${\bf R}(t)={\bf b} + {\bf v}t$,
characterized by the impact-parameter vector ${\bf b}$ and the constant
velocity vector ${\bf v}$ with ${\bf b} \perp {\bf v}$, 
while the rovibrational motion of the target molecule
during the collision is neglected. This implies that a time-dependent
Schr\"odinger equation for an $N$-electron system in a multi-center
Coulomb field (one source moving on a straight-line path, the others fixed
in space) needs to be solved---too complicated 
a problem except for special cases despite ever-growing computational resources.
Accordingly, approximations are in order.

\subsection{Independent atom model}
\label{sec:iam-pcm}
If a molecule were nothing but a collection of atoms with which
the impinging projectile would interact independently,
the net cross sections for capture and ionization  would
be simple sums of atomic net cross sections $\sigma_{x\mid j}^{\rm net}$ for the $j=1,\ldots , M$
atoms that make up the molecule:
\begin{equation}
\sigma_{x\mid {\rm AR}}^{\rm net} = \sum_{j=1}^M \sigma_{x\mid j}^{\rm net} ,
\label{eq:arnet}
\end{equation}
where $x={\rm cap}$ denotes capture and $x={\rm ion}$ ionization.
This prescription is Bragg's additivity rule (AR).
We can picture each atomic cross section as
a circular disk of radius $r_{x\mid j}=\sqrt{\sigma_{x\mid j}^{\rm net}/\pi}$ 
in the impact parameter plane with the atomic center as origin. 
An effective way to represent the cross sectional areas is pixelization,
with the total number of pixels distributed uniformly over each such disk
returning the atomic cross section (when multiplied by the area occupied by
one pixel). 
This suggests an interpretation in which each pixel is associated with 
a scattering event off that atom 
such that the totality of all events is a measure of the net cross section.

In a lateral view, depending on the size and orientation of the
molecule the impinging projectile may encounter $m>1$
pixels along its (straight-line) path. 
The main idea of the IAM-PCM is that
only one instead of $m$ pixels contribute
to the molecular cross section, i.e., multiple scattering events
along a given path are deemed non-existent.
This corresponds to the geometrical interpretation of the IAM-PCM given in our previous
works, namely that the molecular cross section 
is the combined effective area of overlapping atomic cross sections encountered 
by the impinging projectile~\cite{hjl16, hjl18}. 

In order to fully define the model, we have to make a determination
as to where the contributing scattering event happens. 
A classical picture may favor the 
first pixel encountered by the projectile as the one that counts,
whereas the assumption of a statistical distribution seems
more consistent with quantum-mechanical scattering. We
adopt that latter view and assign each pixel along the projectile path
with the weight
factor $1/m$. For the net cross sections this choice is no
different from any other, so long as the sum of all (fractional) contributions
along the path is one. The 
charge-state correlated analysis discussed in Sec.\ref{sec:iem}, 
however, is based on impact-parameter-dependent probabilities
and, as a consequence, is sensitive to that choice.

We can see this in the following way.
The molecular net cross section can be cast into the form
\begin{equation}
\sigma_{x\mid {\rm PCM}}^{\rm net} = \sum_{j=1}^M s_{x\mid j} \sigma_{x\mid j}^{\rm net} ,
\label{eq:pcmnet}
\end{equation}
with weight factors $0 \le s_{x\mid j} \le 1$, which are the
ratios of the fractional contributions to the
total number of pixels within the atomic cross sectional area $\sigma_{x\mid j}^{\rm net}$.
The molecular cross section depends on the kinetic energy of the projectile and,
unlike its AR counterpart (\ref{eq:arnet}), on the orientation of the molecule via
the weight factors. For brevity, these dependences are not indicated in the equations
except further below in Eqs.~(\ref{eq:sigmakl}) and~(\ref{eq:numTCS}).

The first step in a probability-based multiple capture and ionization analysis
is the definition of net electron numbers associated with the molecular net
cross sections. For an ion--atom collision
with the atomic center located at position
${\bf r}_j$ with respect to the center of the molecule, 
the atomic impact parameter is $b_j = |{\bf b} - {\bf r}_j|$, and the atomic
net cross section is obtained by the usual integration over the impact parameter,
exploiting cylindrical symmetry
\begin{equation}
\sigma_{x\mid j}^{\rm net} = 2\pi \int_0^\infty b_j P_{x\mid j}^{\rm net}(b_j) db_j ,
\label{eq:atomnet}
\end{equation}
where the atomic net numbers $P_{x\mid j}^{\rm net}$ are calculated from solving the
Schr\"odinger equation of the ion--atom scattering problem in some approximation (see Sec.~\ref{sec:atomic-theory}
for a short account on our approach).
Inserting Eq.~\ref{eq:atomnet} into Eq.~\ref{eq:pcmnet} yields
\begin{eqnarray}
 \sigma_{x\mid {\rm PCM}}^{\rm net} &=& 
\sum_{j=1}^M s_{x\mid j} \int P_{x\mid j}^{\rm net}(b_j) d^2b_j \\
&\equiv& \int  P_{x\mid {\rm PCM}}^{\rm net}({\bm b}) d^2b
\end{eqnarray}
with
\begin{equation}
P_{x\mid {\rm PCM}}^{\rm net}({\bm b}) = 
	\sum_{j=1}^M s_{x\mid j} P_{x\mid j}^{\rm net}(b_j) .
\label{eq:pmolnet}
\end{equation}

Since the weight factors $s_{x\mid j}$ depend on the fractional contributions
from all pixels 
associated with the atomic cross sections $\sigma_{x\mid j}^{\rm net}$,
so do the net numbers $P_{x\mid {\rm PCM}}^{\rm net}$ and the IEM
probabilities and cross sections derived from them.

\subsection{Independent-electron-model analysis of multiple capture and ionization}
\label{sec:iem}
The probability 
of capturing $k$ and simultaneously ionizing $l$ electrons of an $N$-electron system ($k+l \le N$)
is a multi-dimensional integral of the $N$-particle density $\gamma^N$ of the system taken
at a final time $t_f$ long after the collision~\cite{Luedde03}:
\begin{equation}
P_{kl} = \binom{N}{k+l}\binom{k+l}{l}\int_{P^kI^lT^{N-k-l}}\gamma^N (x_1,\ldots , x_N;t_f) d^4x_1 \dots d^4x_N .
\label{eq:pkl-exact}
\end{equation}
In Eq.~(\ref{eq:pkl-exact}), $x_j$ comprises the space and spin coordinates of the $j{\rm th}$ electron
and
$\int d^4x_j$ is short-hand for 
integration over space and summation over spin.
The spatial integrals are with respect to subspaces $P,T,I=V-P-T$, which
correspond to finding an electron bound to the projectile ($P$), the target ($T$), or
released to the ionization continuum ($I$).

In an IEM analysis in which exchange is neglected (Hartree approximation), the $N$-particle density takes the form~\cite{hjl22}
\begin{equation}
\gamma^N (x_1,\ldots , x_N) = \frac{1}{N^N} \prod_{j=1}^N \gamma^1(x_j) ,
\end{equation}
and the charge-state correlated probabilities of Eq.~(\ref{eq:pkl-exact}) reduce to
\begin{eqnarray}
P_{kl}^{\rm IEM} &=& \binom{N}{k+l}\binom{k+l}{l} \frac{1}{N^N}
 \bigg(\int_P \gamma^1(x) d^4x \bigg)^k
 \bigg(\int_I \gamma^1(x) d^4x \bigg)^l
 \bigg(\int_T \gamma^1(x) d^4x \bigg)^{N-k-l} \nonumber\\
&=& \binom{N}{k+l}\binom{k+l}{l} \frac{1}{N^N} 
(P_{\rm cap}^{\rm net})^k(P_{\rm ion}^{\rm net})^l(P_{\rm tar}^{\rm net})^{N-k-l} ,
\label{eq:pkl1}
\end{eqnarray}
where
\begin{equation}
\int_P \gamma^1(x) d^4x +  \int_I \gamma^1(x) d^4x + \int_T \gamma^1(x) d^4x \equiv 
P_{\rm cap}^{\rm net}+P_{\rm ion}^{\rm net}+P_{\rm tar}^{\rm net} = N .
\end{equation}
If we define average single-particle probabilities by
$P_x= P_x^{\rm net}/N$,
Eq.~(\ref{eq:pkl1}) takes the familiar form~\cite{PhysRevA.49.4556,Schultz_1990}
\begin{equation}
P_{kl}^{\rm IEM}=\binom{N}{k+l}\binom{k+l}{l}  
P_{\rm cap}^kP_{\rm ion}^l(1-P_{\rm cap}-P_{\rm ion})^{N-k-l}  .
\end{equation}
In the context of the IAM-PCM, the single-particle probabilities are obtained from
the molecular net numbers~(\ref{eq:pmolnet}). Hence, the multiple capture and
ionization probabilities depend on 
the kinetic energy of the projectile $E$, the
impact parameter vector ${\bf b} = (b,\phi)$, 
and the Euler angles $\alpha, \theta, \varphi$ which determine the
orientation of the target molecule (we use $z$-$y$-$z$ convention~\cite{Rose95}):
$P_{kl}^{\rm IEM}\equiv P_{kl}^{\rm IEM}(E,b,\phi \mid \alpha,\theta, \varphi)$.
Total cross sections for a given orientation of the molecule are obtained by integration over the
impact parameter vector 
\begin{equation}
\sigma_{kl}^{\rm IEM}(E \mid \alpha,\theta, \varphi) =
\int_0^{2\pi}d \phi \int_0^\infty b db \; P_{kl}^{\rm IEM}(E,b,\phi \mid \alpha,\theta, \varphi) ,
\label{eq:sigmakl}
\end{equation}
and orientation-averaged cross sections are calculated according to
\begin{equation}
\bar \sigma_{kl}^{\rm IEM}(E) = \frac{1}{8\pi^2}\int_0^{2\pi}d \varphi
   \int_0^{\pi}\sin \theta d\theta \int_0^{2\pi}d \alpha \;
\sigma_{kl}^{\rm IEM}(E \mid \alpha,\theta, \varphi) .
\label{eq:numTCS} 
\end{equation}  
The orientation-averaged cross sections can be compared with experimental data obtained for 
randomly oriented molecules.
Since all results discussed in Sec.~\ref{sec:results} are orientation averages, we omit the bar for
ease of notation and denote these cross sections as $\sigma$ (with appropriate sub- and superscripts).

Often, more inclusive cross sections are measured, e.g., for
$q$-fold capture and ionization. They are given as
\begin{eqnarray}
\sigma_q^{{\rm cap}\mid {\rm IEM}} &=& \sum_{l=0}^{N-q} \sigma_{ql}^{\rm IEM} , \\
\sigma_q^{{\rm ion}\mid {\rm IEM}} &=& \sum_{k=0}^{N-q} \sigma_{kq}^{\rm IEM} ,
\label{eq:Pq}
\end{eqnarray}
and they sum up to yield the corresponding net cross sections
\begin{equation}
	\sigma_x^{\rm net} = \sum_{q=1}^N q\, \sigma_q^{x\mid {\rm IEM}} .
  \label{eq:sumrule2} 
\end{equation}

The IEM as laid out here may lead to unphysical higher-order capture since it
has nothing built in to take into account that a given projectile might not be able to accommodate
$k$ electrons.
For protons, double capture ($k=2$) can already be considered an anomaly, possible in principle
but rare in practice since it is associated with the formation of a weakly bound negatively-charged ion.
To avoid the occurrence of anomalous and unphysical outcomes
we modify the IEM analysis as follows:
\begin{eqnarray}
\tilde\sigma_{1q} &=& \sum_{k=1}^{N-q} \; k\;\sigma_{kq}^{\rm IEM} ,
\label{eq:correct1} \\
\tilde\sigma_{k>1 q} &=& 0  ,
\label{eq:correct2}
\end{eqnarray}
i.e., we reinterpret the multiple capture cross sections as contributions
to single capture, as a consequence of which single capture can be identified with net capture.
While this procedure is ad hoc, it appears more physically sound
than the alternatives, which are either taking the multiple capture contributions seriously or simply
ignoring them, and it has been shown to yield reasonabe results
in a number of applications~\cite{Murakami12a, hjl22}.

\subsection{Calculation of the atomic contributions}
\label{sec:atomic-theory}

The formalism of the previous sections explains how capture and ionization
cross sections for a molecular target are obtained from atomic contributions within the IAM.
If the interest is in molecular net cross sections only, any method to compute atomic
net cross sections or even the available experimental data can be used as input.  
For the more detailed multiple capture and ionization cross sections, 
impact-parameter-dependent probabilities, ideally obtained from one
method for all channels, are required. In our work, the nonperturbative basis generator method in its
two-center implementation (TC-BGM) is used to generate these data. Since the
TC-BGM has been described in detail in previous publications~\cite{tcbgm, hjl18},
only a brief account is given here.

We are concerned with an ion--atom collision problem in the semiclassical approximation
and the IEM. This implies that we need to solve a time-dependent Schr\"odinger equation for a
single-particle Hamiltonian of the form ($\hbar=m_e=e=4\pi\epsilon_0=1$)
\begin{equation}
	\hat h (t) = -\frac{1}{2} \nabla^2 + v_T(r) - \frac{Q_P}{r_P(t)} 
\label{eq:hiem}
\end{equation}
for all initially populated orbitals. In Eq.~(\ref{eq:hiem}),
$v_T$ is an atomic target potential, $Q_P$ the charge number of the (bare) projectile
ion ($Q_P=1$ for protons), and $r_P$ the distance between the active (target) electron
and the projectile, which depends on time via the classical projectile trajectory.
As in our previous works~\cite{hjl19, hjl19b, hjl22}, we use atomic target potentials obtained from the
optimized potential method of density functional theory. Exchange is included
exactly while correlation is neglected. Response effects, i.e., time-dependent
changes of $v_T$ over the course of the collision are not taken into account.
The TC-BGM is a coupled-channel method based on bound atomic orbitals on both centers and
a set of pseudostates constructed by repeated application of a regularized potential
operator onto the bound target states. The pseudostates overlap with the continuum and, when
orthogonalized to the bound states, represent the ionization flux at asymptotic times. 

The net numbers in Eq.~(\ref{eq:atomnet})
are obtained by summing up the single-particle probabilities contributing to capture and
ionization of all active orbitals of the target atom. Partially filled subshells are assumed
to be populated statistically, e.g., for oxygen [O$(1s^2 2s^2 2p^4)$]
each of the $2p_{-1}, 2p_0, 2p_{+1}$ 
orbitals is associated with the fractional occupation number $4/3$~\cite{PhysRevA.61.052710}.

\section{Results and Discussion}
\label{sec:results}

We begin the discussion of results in Sec.~\ref{sec:atomic-results} with a look at the net capture and ionization cross sections for proton collisions with atoms which are constituents of biomolecules.
In Sec.~\ref{sec:10el} we turn to the ten-electron systems $\rm Ne, HF, H_2O, NH_3, CH_4$ and discuss net and
charge-state correlated cross sections. We introduce a reduced cross section and show that it follows
a universal curve at high impact energies for proton collisions with $\rm HF, H_2O, NH_3, CH_4$. The reduced
cross section and the associated scaling behaviour also feature prominently in the analysis of 
collisions with biomolecules presented in Sec.~\ref{sec:bio}.
A simple parametrization is shown to work remarkably well over a large impact energy range.

\subsection{Atomic cross sections}
\label{sec:atomic-results}

\begin{figure}
\begin{center}$
\begin{array}{cc}
\resizebox{0.5\textwidth}{!}{\includegraphics{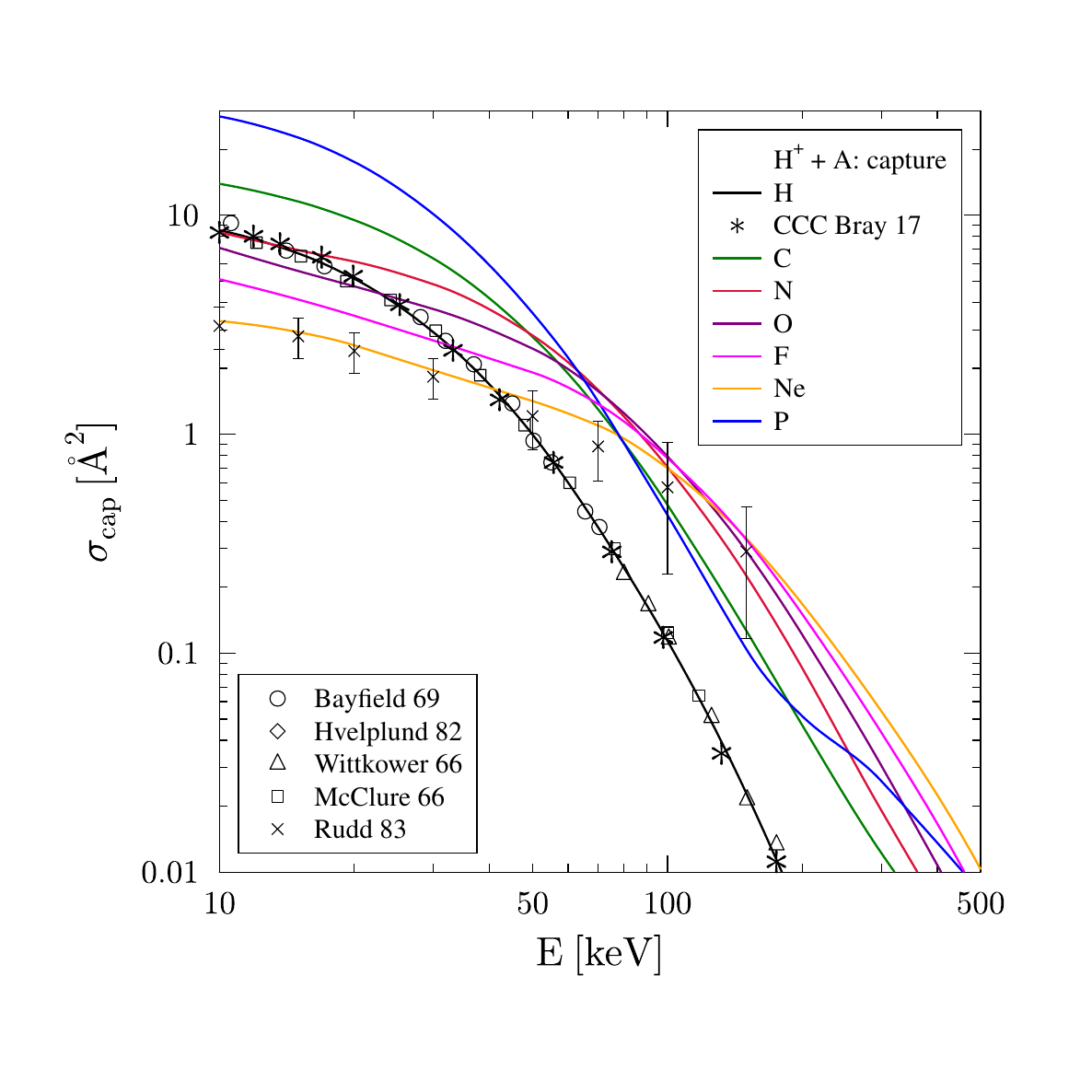}}& \hskip - 0.5 truecm
\resizebox{0.5\textwidth}{!}{\includegraphics{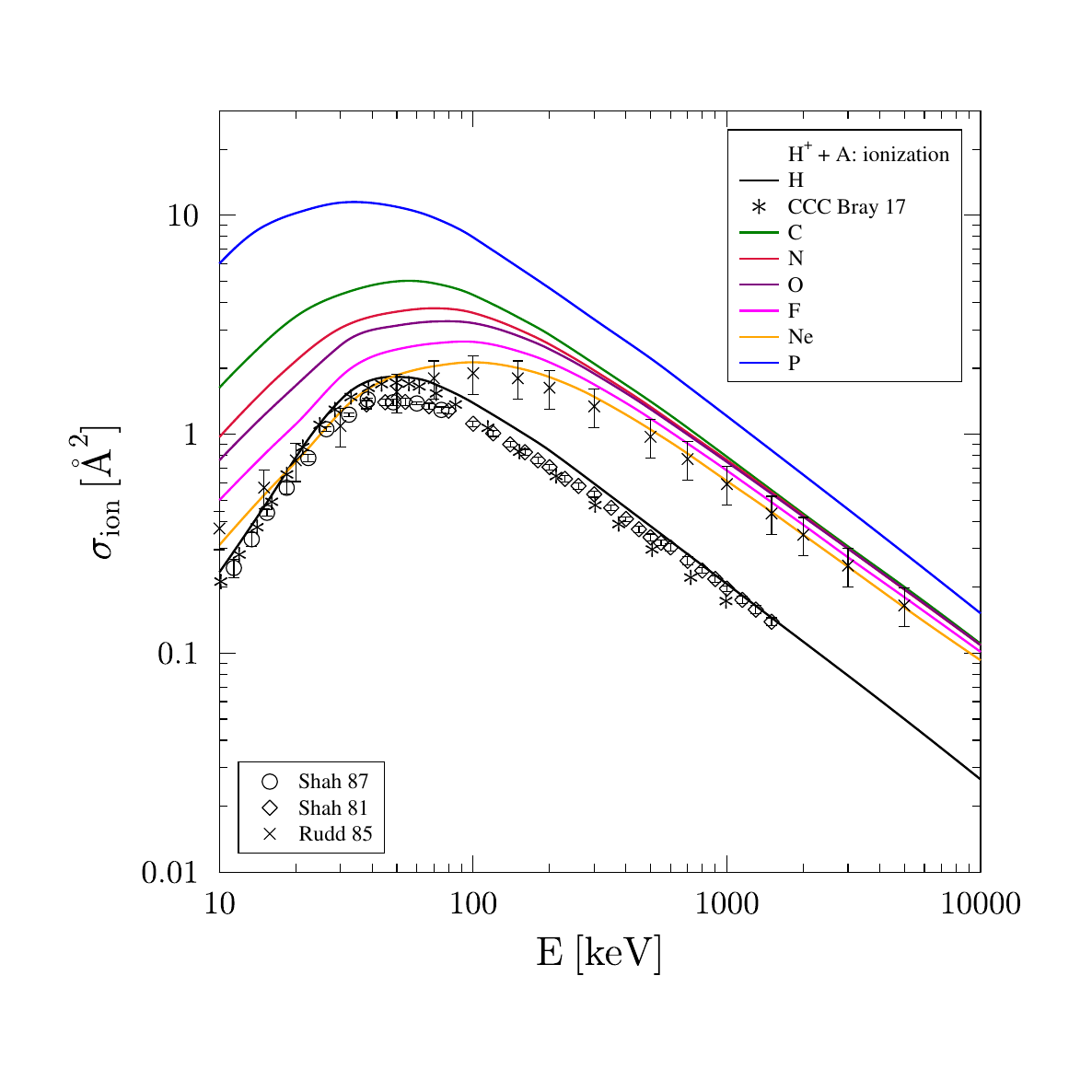}}

\end{array}$
\vskip -0.5 truecm
\caption{
In the left panel the capture cross sections for protons colliding with atoms H, C, N, O, F, Ne, P are shown (solid lines) as calculated with the TC-BGM.
Comparison is made for H atoms with CCC results~\cite{Bray_2017} and experimental data.
The right panel shows corresponding data for net ionization.
References to experimental data can be found in the context of Fig.~1 of Ref.~\cite{hjl18}.
}
\label{fig:Abb1ab}
\end{center}
\end{figure}

We consider the atoms H, C, N, O, F, P, as well as Ne. Figure~\ref{fig:Abb1ab} (left and right panel) shows net capture and ionization by proton impact to indicate their ordering with atomic number. For H the TC-BGM results
are compared with convergent close coupling (CCC) results~\cite{Bray_2017} and with experiments. 
Note that the experimental data for ionization
are lower than the theoretical data at the $20 \%$ level over a wide range. 
The new recommended values for the ionization cross section now adopt the theoretical
values, since many theoretical calculations agree with each other~\cite{Hill_2023}.
For neon targets there are experimental data
available, and our calculations agree well with them, both for capture and net ionization.

For the atoms relevant for biomolecules we find that the capture cross sections at low energies are distributed around the H results:
F and O are lower, N comparable, and C and P higher. At intermediate energies capture from the many-electron atoms is significantly
stronger than for H, and with a change in ordering with atomic number, i.e., crossings occur. At high energies capture from C
is the lowest among the many-electron atoms, while Ne becomes the strongest. For P we observe first a decrease, and then
$L$-shell capture sets in to boost the cross section.

For ionization we find that Ne has the lowest cross section among the many-electron atoms over the entire range.
Among the biologically relevant atoms the ordering is F, O, N, C, P. At high energies net ionization is dominated by
the single-electron process and the parallel behaviour can be understood on the basis of the Bethe-Born model.
In this limit the cross sections for the second-row atoms $\rm C \to F$ become approximately equal, and about four
times as big as for hydrogen. The F atom is ionizing more weakly (at the $10 \%$ level), and P exceeds the other atoms
by $50 \%$. We compared our atomic F cross section at high energies to the positron-fluorine cross section of
Ref.~\cite{Mori_2024} scaled to corresponding proton collision energies, and found that they are quite similar in this regime.
The present atomic cross sections are calculated in a spherical-atom approach (cf. Sec.~\ref{sec:atomic-theory}), and in the high-energy limit they follow the
Bethe-Born behaviour~\cite{PhysRevA.100.062702} as calculated from the Hartree-Fock orbitals.

\subsection{Ten-electron systems}
\label{sec:10el}

\begin{figure}
\begin{center}$
\begin{array}{cc}
\resizebox{0.5\textwidth}{!}{\includegraphics{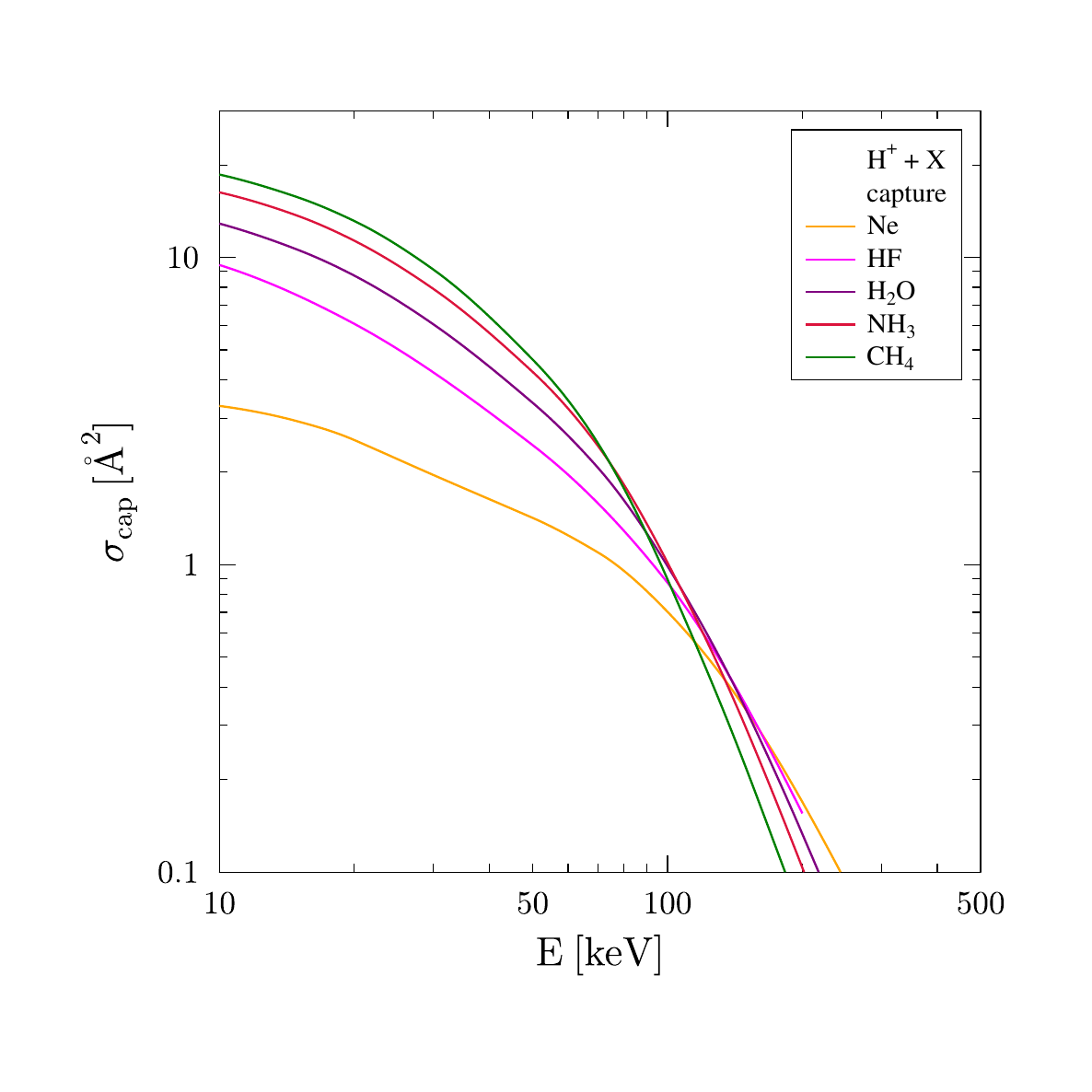}}&  \hskip -0.5 truecm
\resizebox{0.5\textwidth}{!}{\includegraphics{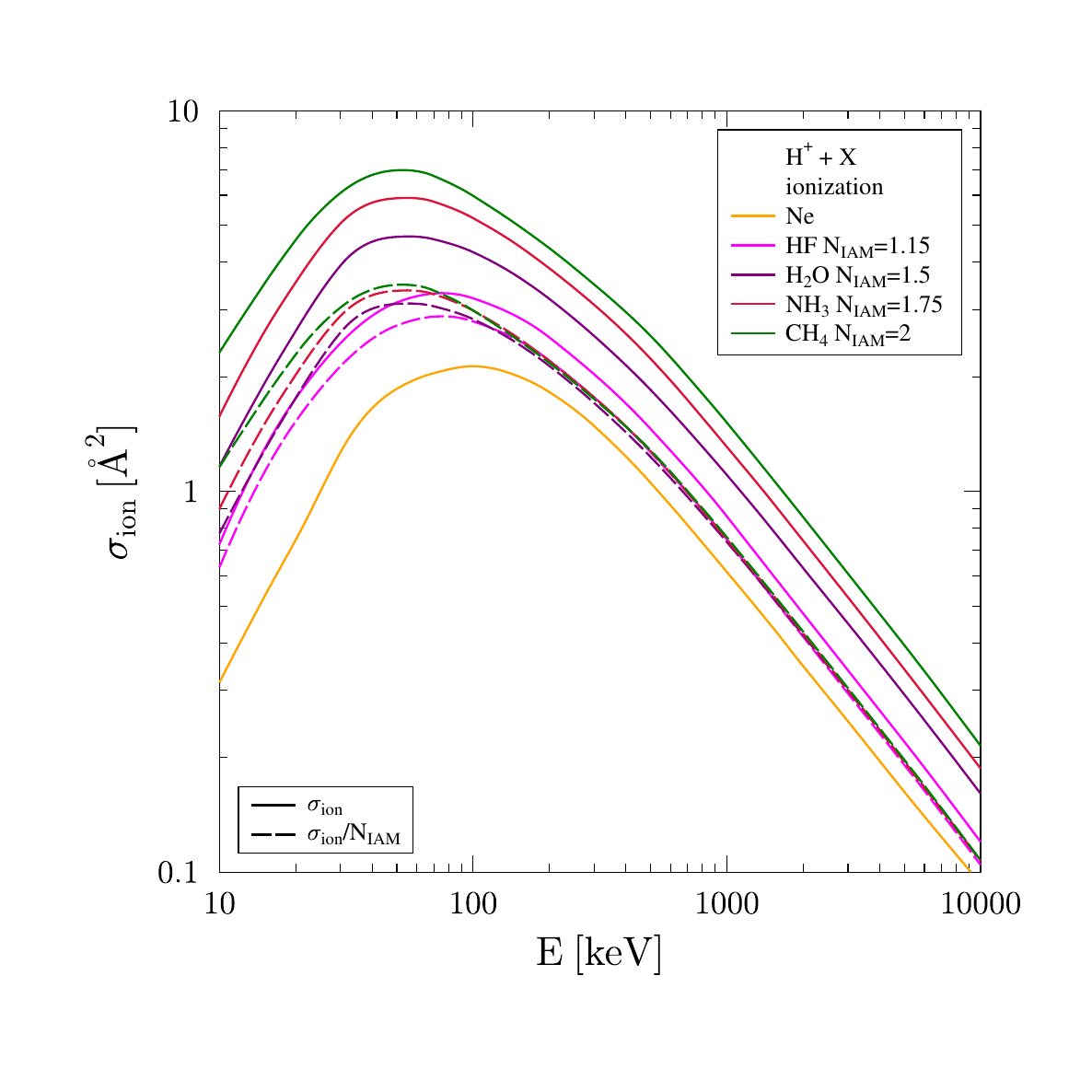}}
\end{array}$
\vskip -0.5 truecm
\caption{
In the left panel the capture cross sections for protons colliding with ten-electron systems 
$\rm Ne, HF, H_2O, NH_3, CH_4$ are shown (solid lines) as calculated with the IAM-PCM based on
the atomic cross sections shown in the left panel of Fig.~\ref{fig:Abb1ab}.
The right panel shows corresponding data for net ionization as solid lines. The dashed lines show the reduced cross section 
Eq.~(\ref{eq:BB2}), and merge into a common curve at impact energies above 100~keV.
}
\label{fig:Abb2ab}
\end{center}
\end{figure}

In Fig.~\ref{fig:Abb2ab} (left and right panel) we present orientation-averaged IAM-PCM net capture and net electron production  cross sections for the sequence $\rm Ne, HF, H_2O, NH_3, CH_4$.
For capture at low collision energies the ordering of the cross sections follows from a combination of which lead atom (C, N, O, F)
is involved and the number of H atoms in the molecule. Similarly to the atomic cross sections of Fig.~\ref{fig:Abb1ab} crossings occur at intermediate energies.

For the electron production at high energies the Bethe-Born limit applies for the atomic
cross sections. Ionization is dominated by the outermost $2p$ orbital.
Based on our calculations, we observe a scaling behaviour for the net cross section, which can be described
by a single parameter $N_{\rm IAM}$ for each molecule:  
\begin{equation}
N_{\rm IAM} =0.25 \, n_{\rm H} + n_{\rm C} +n_{\rm N} +n_{\rm O} + 0.9 \, n_{\rm F}+1.5 \, n_{\rm P} ,
\label{eq:BB1}
\end{equation}
where $n_{\rm H}, n_{\rm C}$, {\it etc.}, are the numbers of constituent atoms for the given molecule. Phosphorous is a constituent of some biomolecules discussed further below.
The net cross section for a given molecule $X$ with composition determined by these numbers
when divided by $N_{\rm IAM}$, i.e.,
\begin{equation}
	\sigma_{{\rm ion}\mid X}^{\rm net }/N_{\rm IAM}
\label{eq:BB2}
\end{equation}
is shown to follow a universal curve at high energies as indicated by dashed lines in the right panel of Fig.~\ref{fig:Abb2ab}.
The coefficients appearing in Eq.~(\ref{eq:BB1}) follow from the atomic cross sections shown in the corresponding Fig.~\ref{fig:Abb1ab}.
We call Eq.~(\ref{eq:BB2}) the reduced IAM cross section.

\begin{figure}
\begin{center}$
\begin{array}{cc}
\resizebox{0.5\textwidth}{!}{\includegraphics{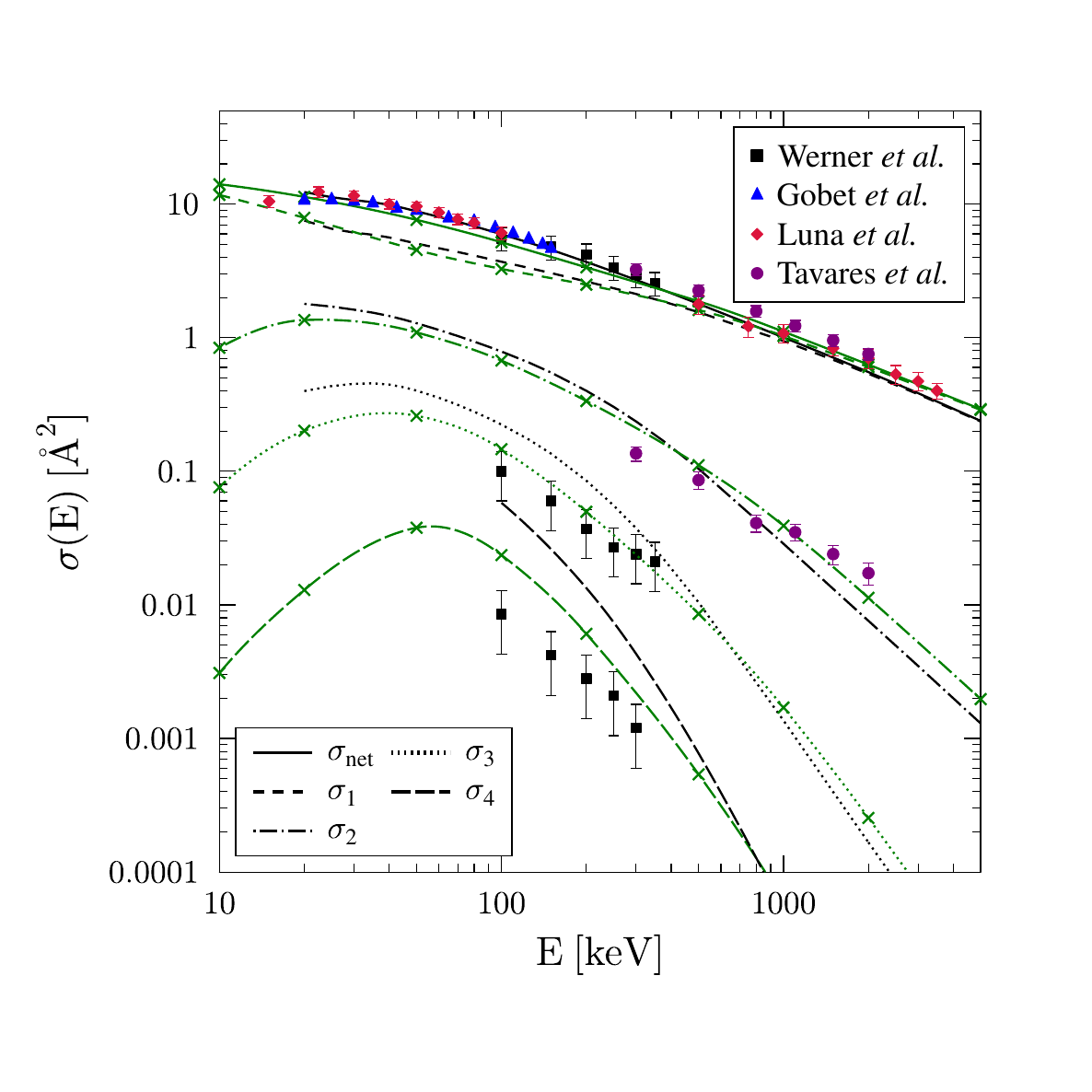}}&  \hskip - 0.5 truecm
\resizebox{0.5\textwidth}{!}{\includegraphics{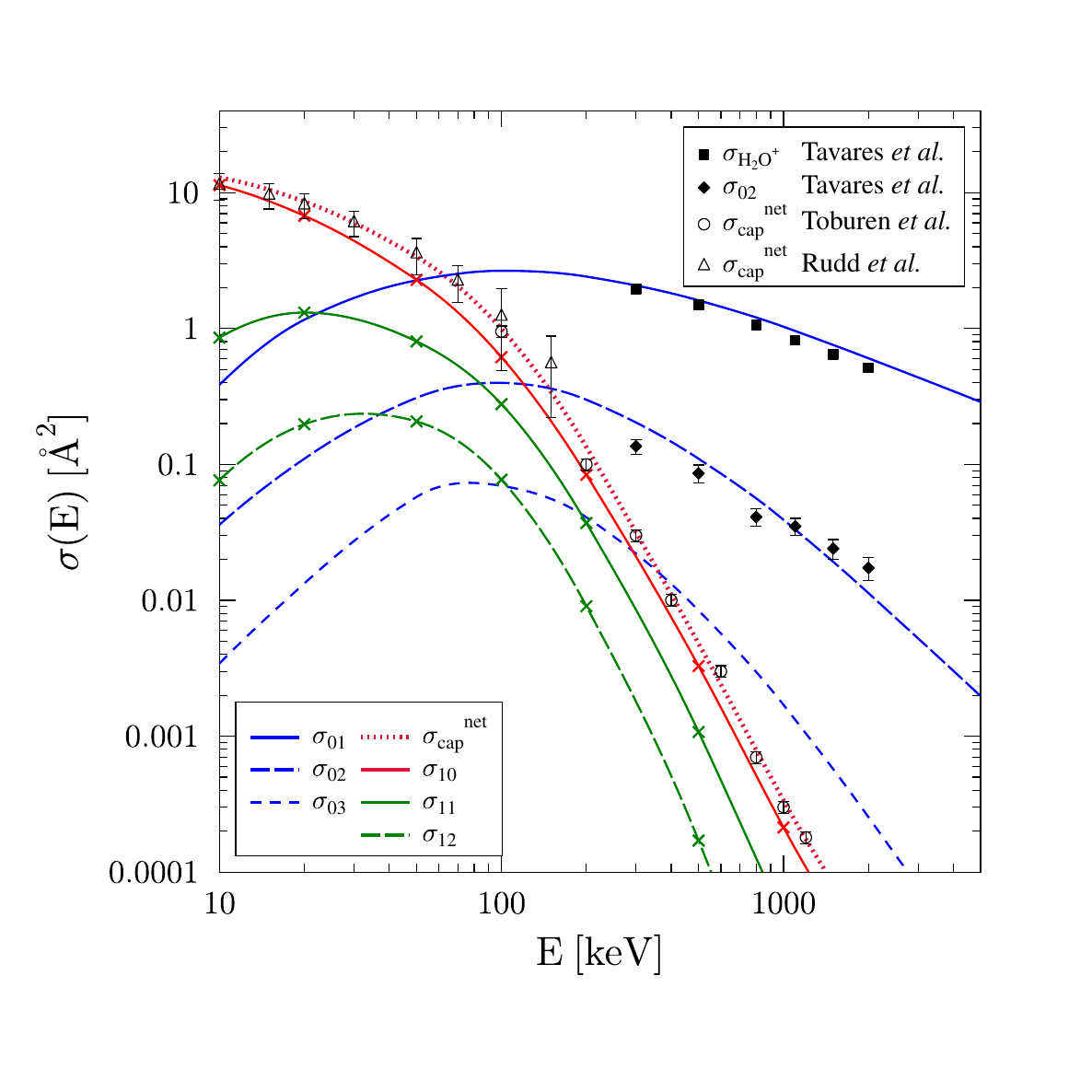}}
\end{array}$
\caption{Charge-state correlated cross sections for proton-$\rm H_2O$ collisions. Left panel:
electron removal, i.e., production of charge states $q^+$ in comparison with experiment.
Solid lines: net cross section, broken lines: $\sigma_q$ with $\rm q=1,2,3,4$. Black: MO-IEM from Ref.~\cite{Kirchner13};
green lines with crosses: IAM-PCM from Ref.~\cite{hjl22}. Right panel: 
separated cross sections $\sigma_{k l}$, in blue: pure ionization, $k=0, l=1,2,3$, in red: pure
capture, $k=1, l=0$, solid green is transfer ionization, $k=1,l=1$; dashed green, $k=1,l=2$
transfer ionization with two electrons in the continuum. The pure capture and transfer ionization cross sections are
obtained from Eq.~(\ref{eq:correct1}). Dotted red line: total capture
is compared with experimental data of Toburen and Rudd, respectively.
Experimental data are from Refs.~\cite{Werner95, PhysRevA.92.032714}
and references 
that can be found in the context of Figs.~2 and~3 of Ref.~\cite{hjl22}.
The left and right panels of the present figure are modifications of Fig.~2(a) and Fig.~3 from Ref.~\cite{hjl22}.
}
\vskip -0.5 truecm
\label{fig:Abb3ab}
\end{center}
\end{figure}

In the left panel of Fig.~\ref{fig:Abb3ab} we show the cross sections for the production of $\rm H_2O^{q+}$ ions in proton--water molecule collisions due to
capture and ionization for $q=1,\ldots ,4$, and how they make up the net electron removal cross section.
At high energies the net cross section is dominated by the $q=1$ channel, but between 20 and 200 keV impact energy
the $q=2$ channel makes a clear contribution to the net cross section (it is weighted by a factor of 2).
Results from two theoretical models are shown, namely the MO-IEM 
of Ref.~\cite{Kirchner13} (black lines), and the IAM-PCM (green lines with crosses).
The experimental data for $q$-fold electron removal are derived from measuring the charged fragments.
For a detailed discussion see Ref.~\cite{hjl22}.

The right panel of Fig.~\ref{fig:Abb3ab} presents the separated capture and ionization contributions as calculated in the IAM-PCM. At higher collision energies
the data agree very well for direct single and double ionization. At low energies transfer ionization ($\sigma_{11}$) is stronger
than pure single ionization  ($\sigma_{01}$), which is a result of the interplay
of capture and ionization probabilities at small impact parameters.
The presence of transfer ionization at intermediate energies leads to the conclusion that net capture at these energies
is larger than pure single capture ($\sigma_{10}$).

\begin{figure}
\begin{center}$
\begin{array}{ccc}
\resizebox{0.35\textwidth}{!}{\includegraphics{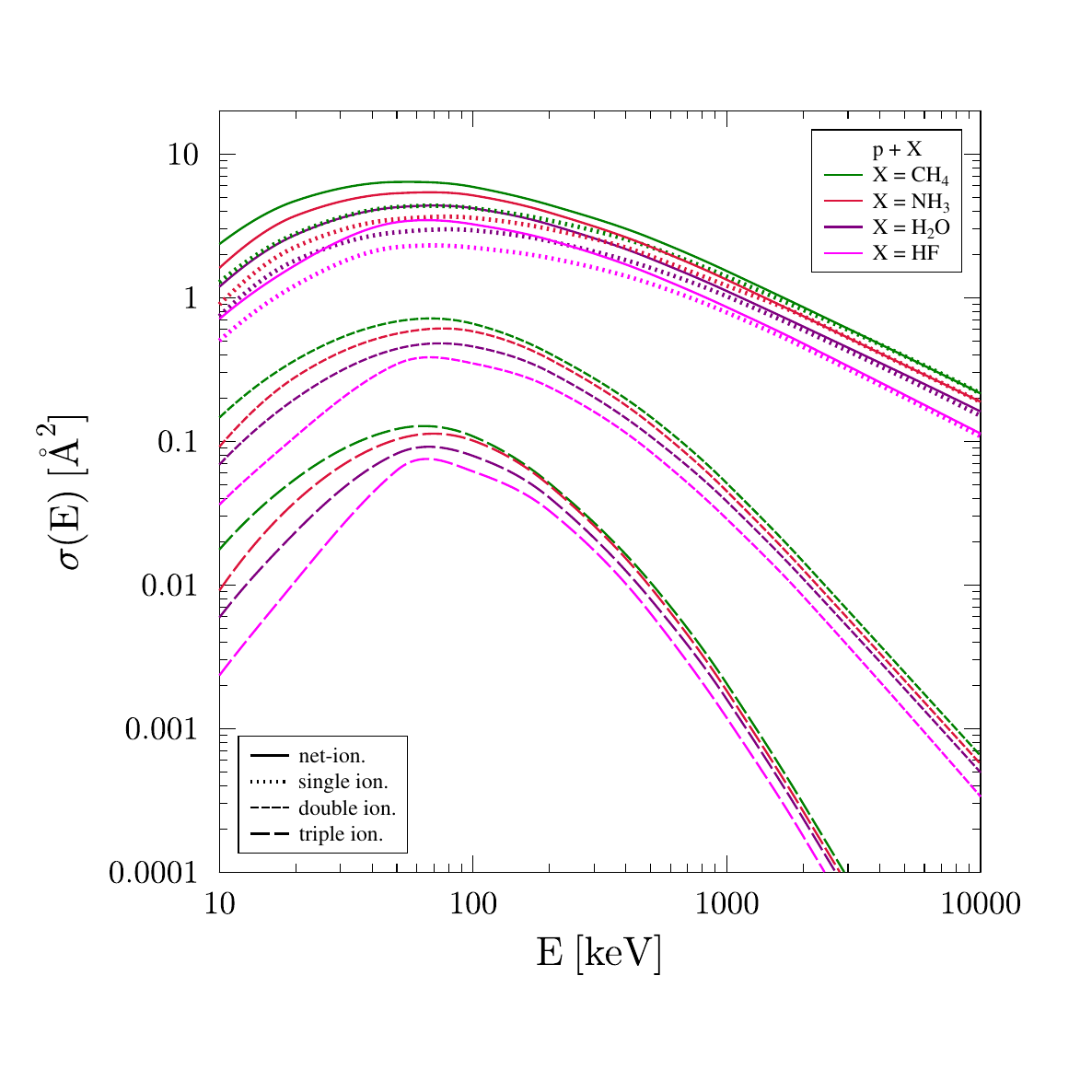}}& \hskip -0.5 truecm
\resizebox{0.35\textwidth}{!}{\includegraphics{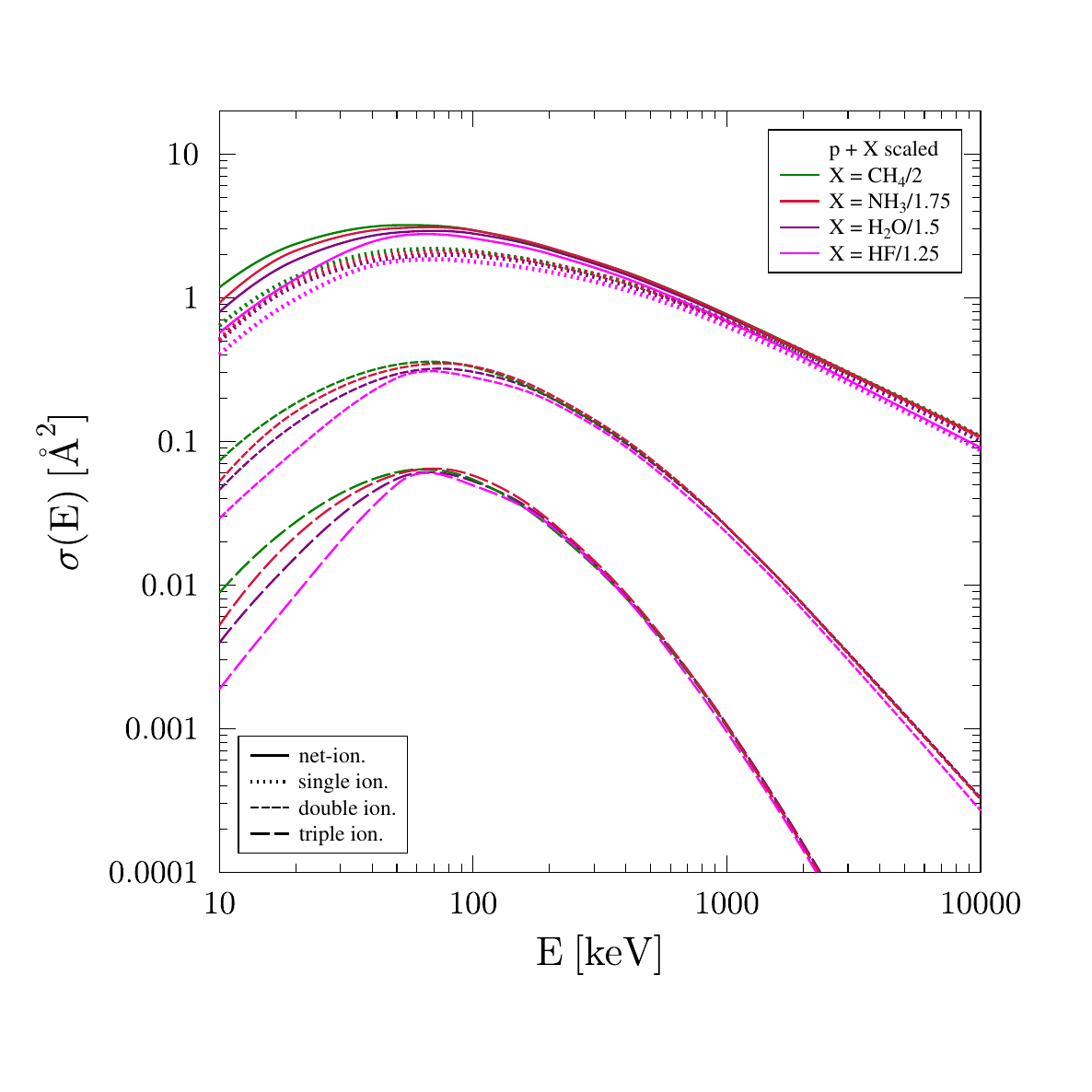}}&  \hskip -0.5 truecm
\resizebox{0.35\textwidth}{!}{\includegraphics{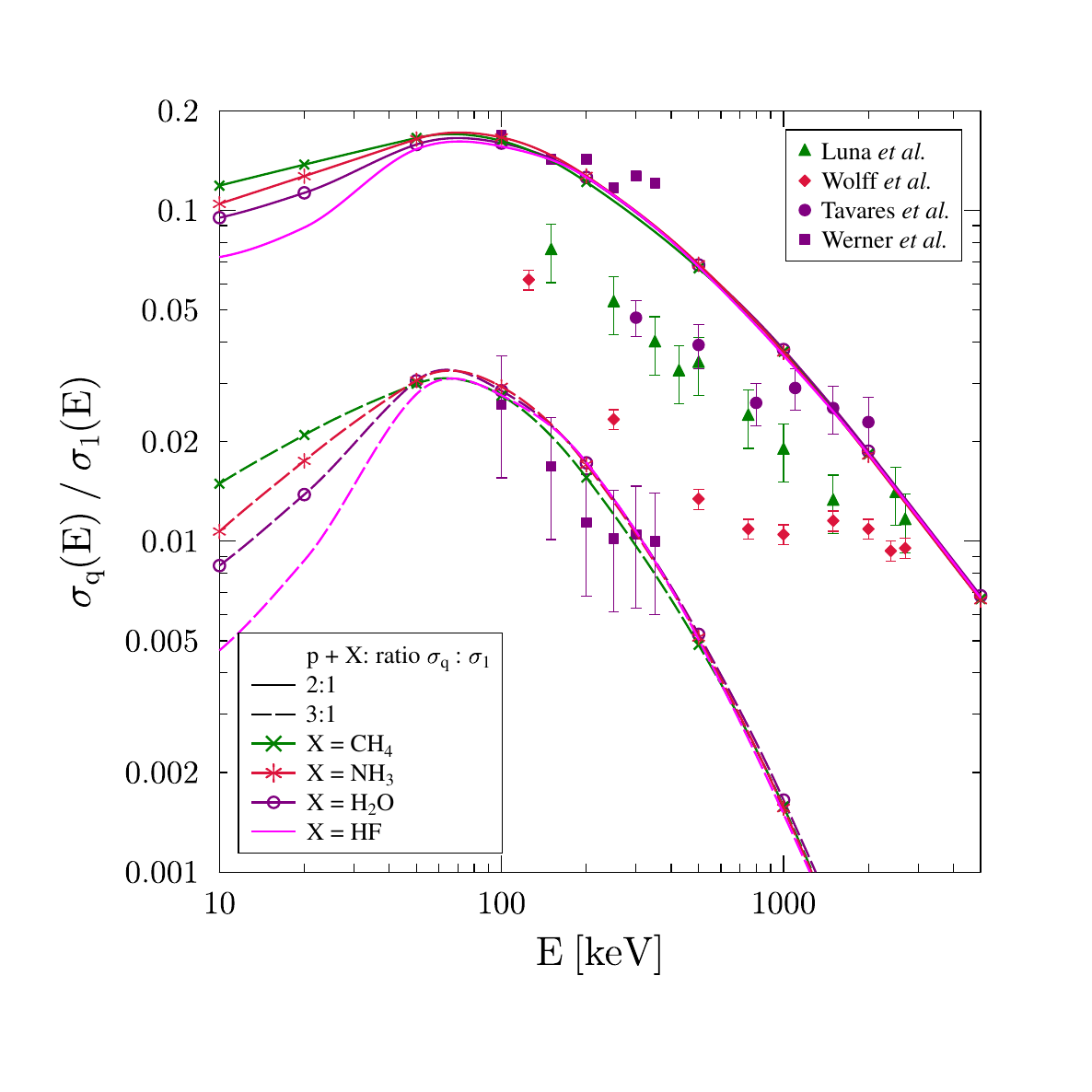}
}
\end{array}$
\vskip -0.5 truecm
\caption{Left panel: IAM-PCM cross sections for net electron production, and $q$-fold electron production $\sigma_{0 \, q}+\sigma_{1 \, q}$ in proton-X collisions
where $\rm X = HF, H_2O, NH_3, CH_4$.
Middle panel: same as left panel, but obtained with the IAM scaling model, i.e., Eq.~(\ref{eq:BB2}) applied to net and $q$-fold electron production. 
Right panel: comparison of ratios for $q$-fold
electron production: $\sigma_q/\sigma_1$ for $q=2,3$ with experimental results from
Refs.~\cite{Werner95,PhysRevA.92.032714,Luna19,Wolff20}.
}
\label{fig:Abb4ac}
\end{center}
\end{figure}

We now investigate the scaling behaviour of charge-state correlated cross sections due to proton impact for the sequence $\rm HF, H_2O, NH_3, CH_4$.
Figure~\ref{fig:Abb4ac} shows the calculated cross sections in the left panel for net electron production, and
the electron production separated into the $q=1,2,3$ channels. The ordering in $q$-fold electron production is maintained as compared to 
$\sigma_1$ or the net production. A small shift in collision energy can be seen for the location of the maximum with increasing $q$.

The middle panel shows that the scaling works not only at high collision energies, but down to the maximum at about $70 \, \rm keV$
proton impact energy. The calculated ionization cross sections do not include post-collision effects, such as autoionization, which may
play a role at higher energies.

Finally, in the right panel the ratios $\sigma_q/\sigma_1$ are compared with available experimental data.
The solid lines show that the universality (scaling implied by Eq.~(\ref{eq:BB2})) is working down to collision energies of $50 \rm \, keV$.
Note that these ratios do not depend on the scaling parameter $N_{\rm IAM}$ of Eq.~(\ref{eq:BB2}), since it cancels. 
For $\rm H_2O$ the data of Werner~{\it et al.}~\cite{Werner95} shown as solid squares are found to be in good agreement 
with the theoretical model. The results of Tavares~{\it et al.}~\cite{PhysRevA.92.032714} show a smaller value for the ratio
by about a factor of two. A strong reason for trusting the work of Werner~{\it et al.}  is that they had a complete
coincidence count, which is not obvious for the data of Tavares~{\it et al.}
The rise for energies of  1\,000 keV and higher was interpreted as coming 
from autoionization contributions in Ref.~\cite{PhysRevA.92.032714}, but their data do not connect with those of Werner~{\it et al.} where they overlap, suggesting that undetected multiple proton events might be, at least in part, to blame for the disagreement.

For methane ($\rm CH_4$) the ratios $\sigma_2/\sigma_1$ of Luna~{\it et al.}~\cite{Luna19} agree with the water molecule results
of the same group (Tavares~{\it et al.}), but for ammonia ($\rm NH_3$) the data of Wolff~{\it et al.}~\cite{Wolff20} start out in agreement
at about 125 keV impact energy, but then fall off more steeply and
are lower by a factor of two at intermediate energies. Again, at about 1\,000 keV proton energy the apparent rise of the ratio
was interpreted as the onset of autoionizing contributions~\cite{Wolff20}. A similar onset appears to be seen at a higher
collision energy for methane.

The present results for the ratios $\sigma_q/\sigma_1$ are demonstrated to be nearly universal for the four ten-electron molecules
at energies of 100 keV and higher. This observation is independent of the details of the IAM. We believe strongly that in the future
more complete fragment observations (including all coincidences, particularly for multiple proton production) are 
warranted for methane and ammonia targets to settle the question as to whether
the step-like behaviour of the ratios  $\sigma_2/\sigma_1$ at 1\,000 keV impact energy for the water
and ammonia data (which for methane appears only at 2\,000 keV) are due to autoionzation contributions, as argued
by the experimentalists, or caused by the lack of detecting multiple proton events.
These events were explicitly counted in the 
coincidence work of Werner~{\it et al.}
A resolution of the present discrepancy would represent an important milestone in the field.

\subsection{Complex biomolecules}
\label{sec:bio}

The calculation of proton collisions with biomolecules requires molecular structure files. 
We have taken them from the MolView website~\cite{molview}.
Our IAM-PCM results are presented as follows: We begin with the pyrimidines and purines, and then
follow up with the amino acids. These are composed of atoms H, C, N, O.
Nucleotides contain in addition a phosphate group, and are larger in size.
We demonstrate how effective the reduced cross section approach is, and
how the dependence on the parameter $N_{\rm IAM}$ characterizes the electron production cross sections.
We then describe how one can model the net ionization cross sections for
the four molecular groups with the help of a Bethe form generalized with the help of
an effective charge parameter.

\begin{figure}
\begin{center}$
\begin{array}{cc}
\resizebox{0.5\textwidth}{!}{\includegraphics{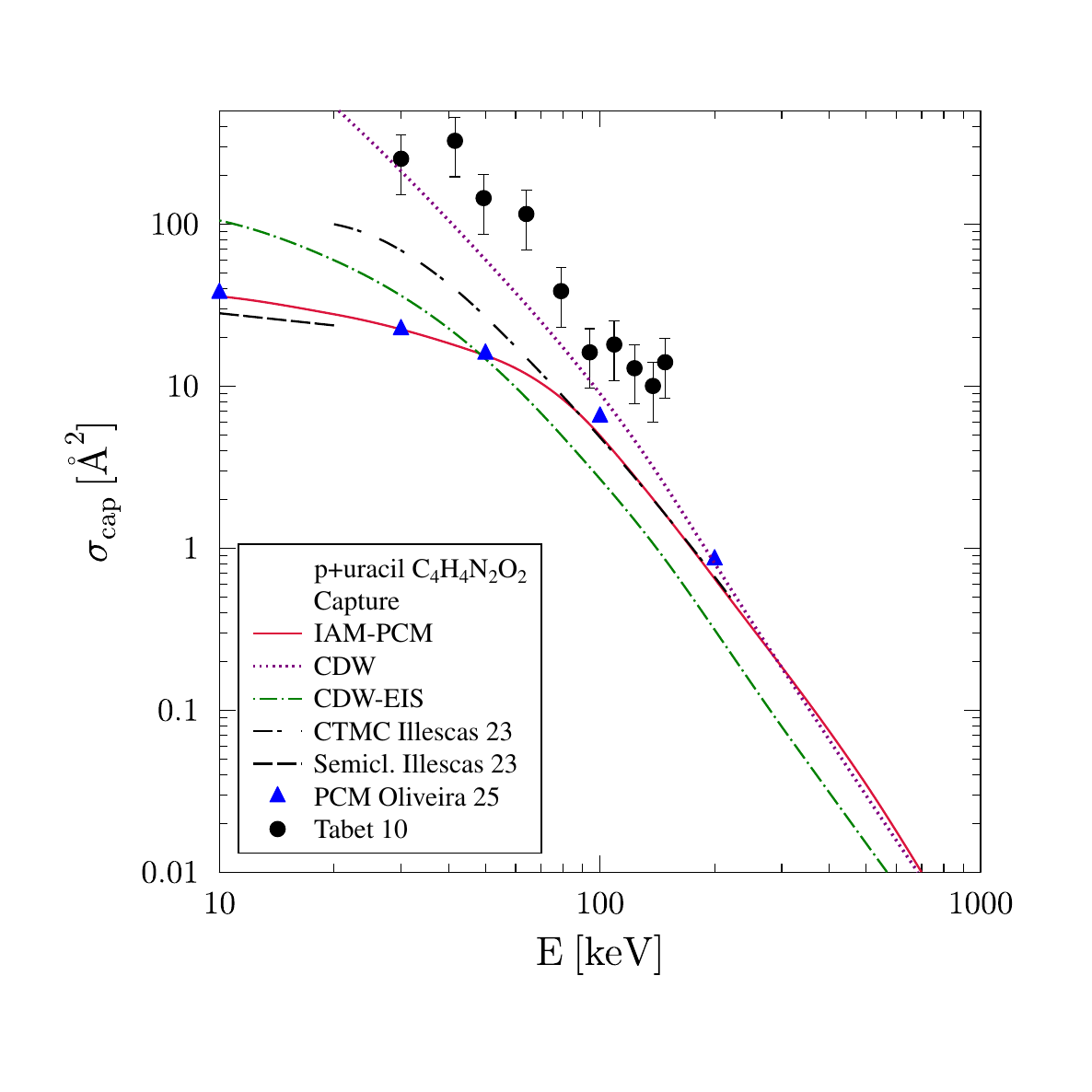}}&  \hskip - 0.7 truecm
\resizebox{0.5\textwidth}{!}{\includegraphics{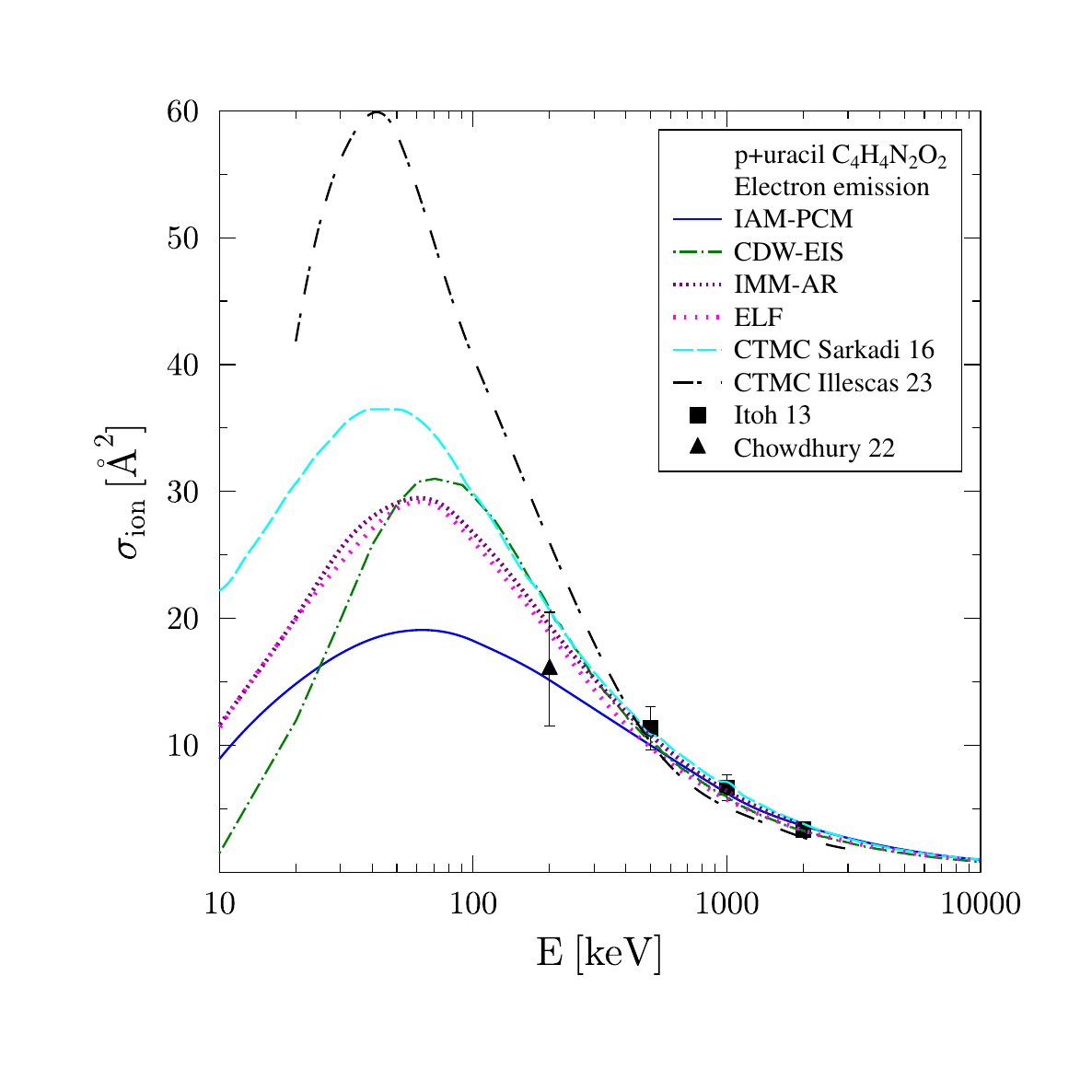}
}
\end{array}$
\vskip -0.5 truecm
\caption{Left panel: net capture in proton-uracil collisions;
solid line (red): present IAM-PCM; triangles (blue): corrected PCM~\cite{Bernal2024}
(private communication);
dash-dotted (green) line: CDW-EIS~\cite{Champion_2012}; dotted (magenta) line CDW~\cite{Belkic_2021}. 
Black solid circles: experiments of 
Tabet et al.~\cite{PhysRevA.82.022703}.
Right panel: net electron emission: solid (blue) line: present IAM-PCM; dashed green line: CDW-EIS~\cite{Champion_2013},
pink dotted line: (ELF)~\cite{Vera_2013}, magenta dotted line: (IMM-AR)~\cite{Paredes15}.
Experimental data: squares~\cite{Itoh13}, triangle~\cite{Chowdhury_2022}.
}
\label{fig:Abb5ab}
\end{center}
\end{figure}

In Fig.~\ref{fig:Abb5ab} we show results for net electron capture and net electron emission in proton collisions with uracil ($\rm C_4 H_4 N_2 O_2$).
For net capture (left panel) we note that the disagreement reported by de Oliveira \textit{et al.}~\cite{Bernal2024} has been resolved and that the data of the two
IAM-PCM calculations agree perfectly at low energies. The data of Ref.~\cite{Bernal2024} are a bit higher where the cross section decreases due
to a difference in atomic cross sections. The experimental data are higher by a large factor.
The CDW-EIS results are lower as compared to IAM-PCM by a factor of two at higher energies, but then rise more steeply when the
cross section increases. 
The recent work of Ref.~\cite{Clara_2023} includes two approaches: a low-energy semiclassical
MO calculation shown for 10--20 keV impact energy and a CTMC calculation. The MO result is close to the IAM-PCM data and provides some
validation, since it should be reliable there. The classical-trajectory calculation from Ref.~\cite{Clara_2023}, on the other hand, 
agrees with IAM-PCM only at energies from about 100 keV on, and reaches higher values at low proton energies.

For net electron emission, shown in the right panel of Fig.~\ref{fig:Abb5ab}, we find a lower and broader shape in the 
cross section at low to intermediate energies than the CDW-EIS results from Ref.~\cite{Champion_2013}. At high energies the theoretical results merge
and agree well with the experimental data of Ref.~\cite{Itoh13}. The data point of at 200 keV impact energy from Ref.~\cite{Chowdhury_2022}
appears to be supportive of the IAM-PCM result as compared to CDW-EIS.
Two CTMC calculations are shown. They disagree with each other by a factor of two near the maximum with the result of Ref.~\cite{Clara_2023} being larger than
that of Ref.~\cite{Sarkadi_2016}. 
We do not show the experimental net ionization data for uracil from Ref.~\cite{PhysRevA.82.022703}, since they appear to be an order of magnitude too high.
This has been discussed, e.g., in Ref.~\cite{hjl16}.

We note that proton collisions with adenine  ($\rm C_5H_5N_5$) lead to very similar results as uracil. 
Given the importance of this purine nucleotide base (it is found in DNA and RNA) there are a number of calculations available, and also
some experimental data. Results analogous to those shown in Fig.~\ref{fig:Abb5ab} can be found in Figs.~3 and 4 of Ref.~\cite{hjl19b}.

\begin{figure}
\begin{center}$
\begin{array}{cc}
\resizebox{0.5\textwidth}{!}{\includegraphics{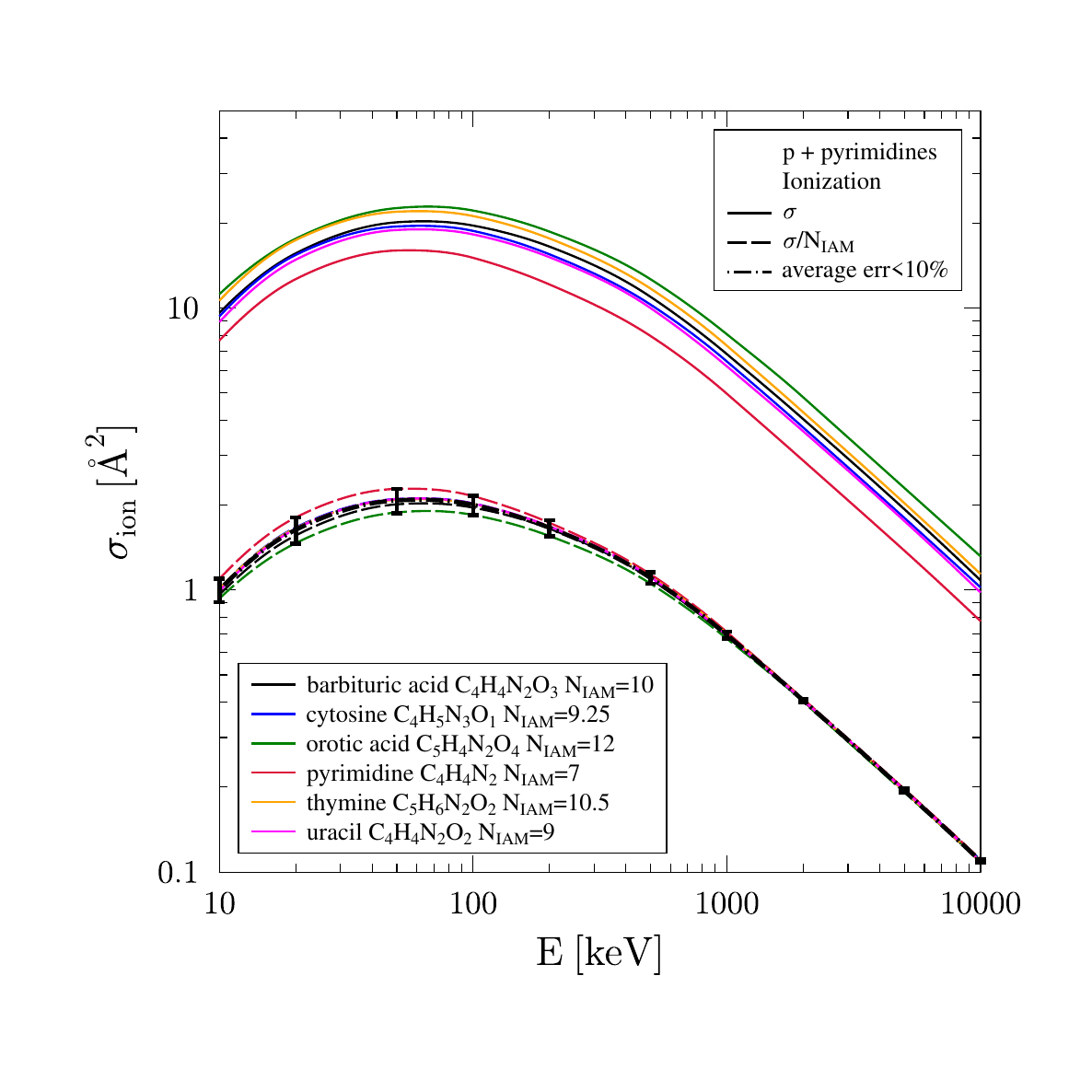}}&  \hskip - .5 truecm
\resizebox{0.5\textwidth}{!}{\includegraphics{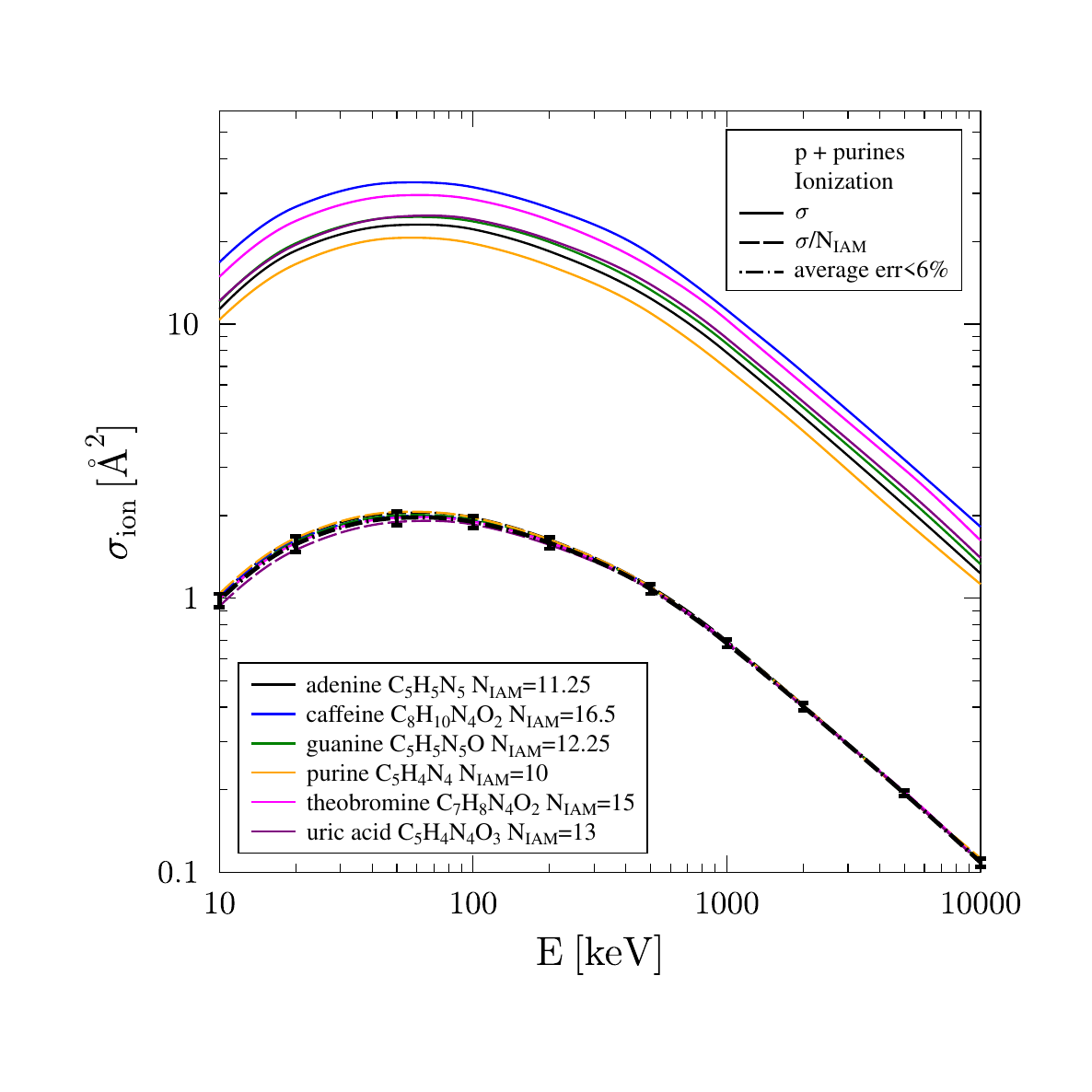}
}
\end{array}$
\vskip -0.5 truecm
\caption{Net ionization of pyrimidines (left panel) and purines (right panel) by proton impact. Note the range of parameter values  $7 \le N_{\rm IAM} \le 12$
for the pyrimidines, and  $10 \le N_{\rm IAM} \le 16.5$ for the purines.}
\label{fig:Abb6ab}
\end{center}
\end{figure}

We proceed with a discussion of net electron production cross sections for four groups of molecules: pyrimidines, purins,
amino acids, and nucleotides.
In the left panel of Fig.~\ref{fig:Abb6ab} we present data for the pyrimidine group of molecules. 
The net electron emission cross sections for the individual molecules are shown
in the top portion of the figure and they have very similar shapes. Orotic acid ($\rm C_5H_4N_2O_4$, shown in green) has the highest value of $N_{\rm IAM}=12$
in this group and has the largest cross section. Pyrimidine on the other hand, ($\rm C_4H_4N_2$, shown in red) has the lowest value ($N_{\rm IAM}=7$), 
and is at the bottom of the group of cross sections. The ratio of the top to the bottom curve is about 1.5 at the lowest energies and 1.7 at the high-energy regime. 

The method of reduced cross sections developed in Sec.~\ref{sec:10el} allows to summarize the results for the entire group.
 The dashed curves shown in the lower part of the left panel of Fig.~\ref{fig:Abb6ab} demonstrates that based on Eq.~(\ref{eq:BB2}) the reduced cross sections
fall within an error band with at most ten percent deviation.

In the right panel of Fig.~\ref{fig:Abb6ab} we show analogous results for the purines, which are based on the same atoms as the pyrimidines,
but are larger, as indicated by the range of $N_{\rm IAM}$ values. Here caffeine ($\rm C_8H_{10}N_4O_2$, shown in blue)
is represented by the top curve, and purine ($\rm C_5H_4N_4$, shown in yellow) is at the bottom.
We find that
the reduced cross sections coalesce even better than for the pyrimidines shown in the left panel.

Amino acids are a class of organic compounds with at least one carboxyl group ($\rm COOH$) and one amino group ($\rm NH_2$).
Biorelevant are the proteinogenic amino acids, which are elements of proteins. 
With the exception of glycine, amino acids have chiral symmetry. 
As proteinogenic amino acids, they only occur as L enantiomers. 

\begin{figure}
\begin{center}$
\begin{array}{cc}
\resizebox{0.5\textwidth}{!}{\includegraphics{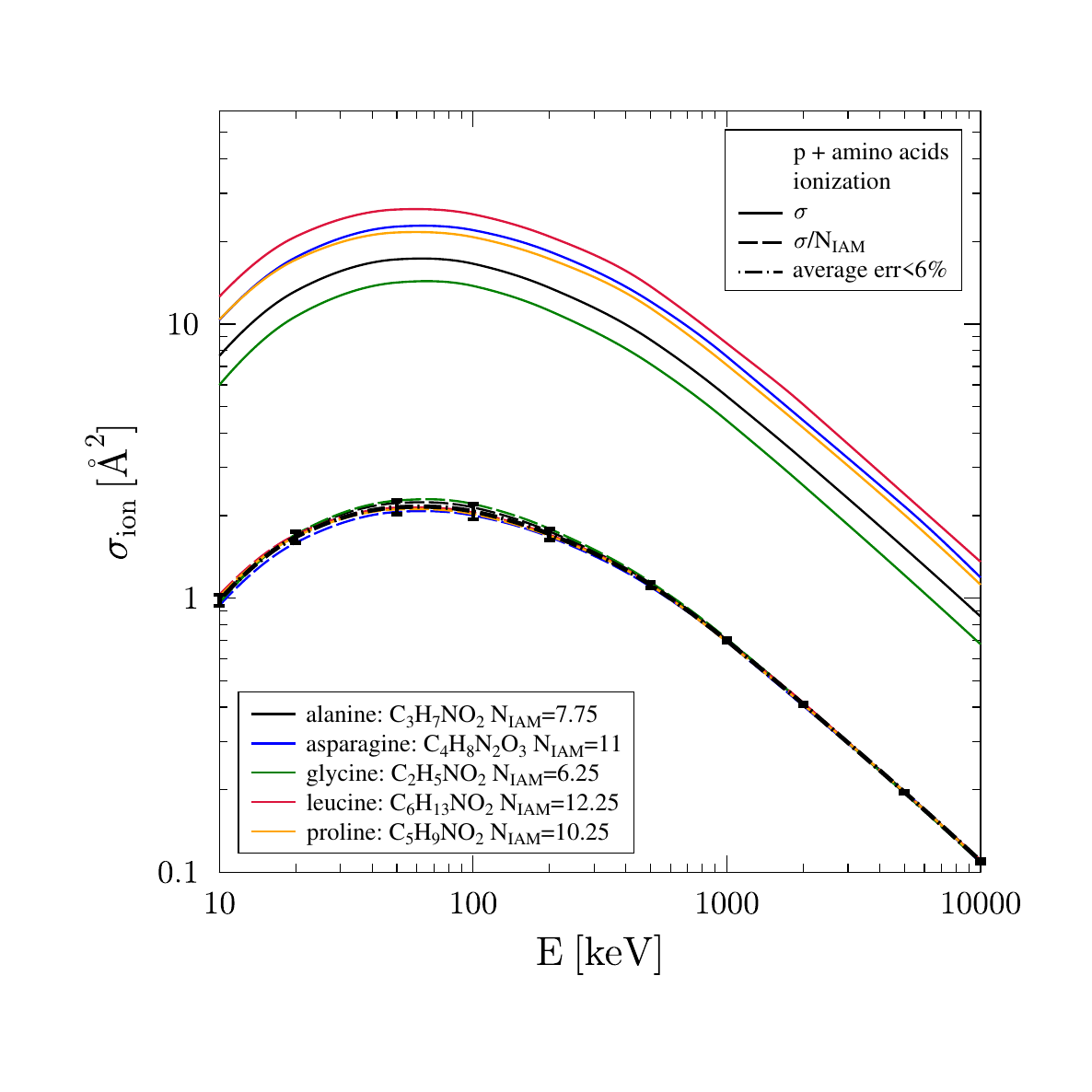}}&  \hskip - 0.5 truecm
\resizebox{0.5\textwidth}{!}{\includegraphics{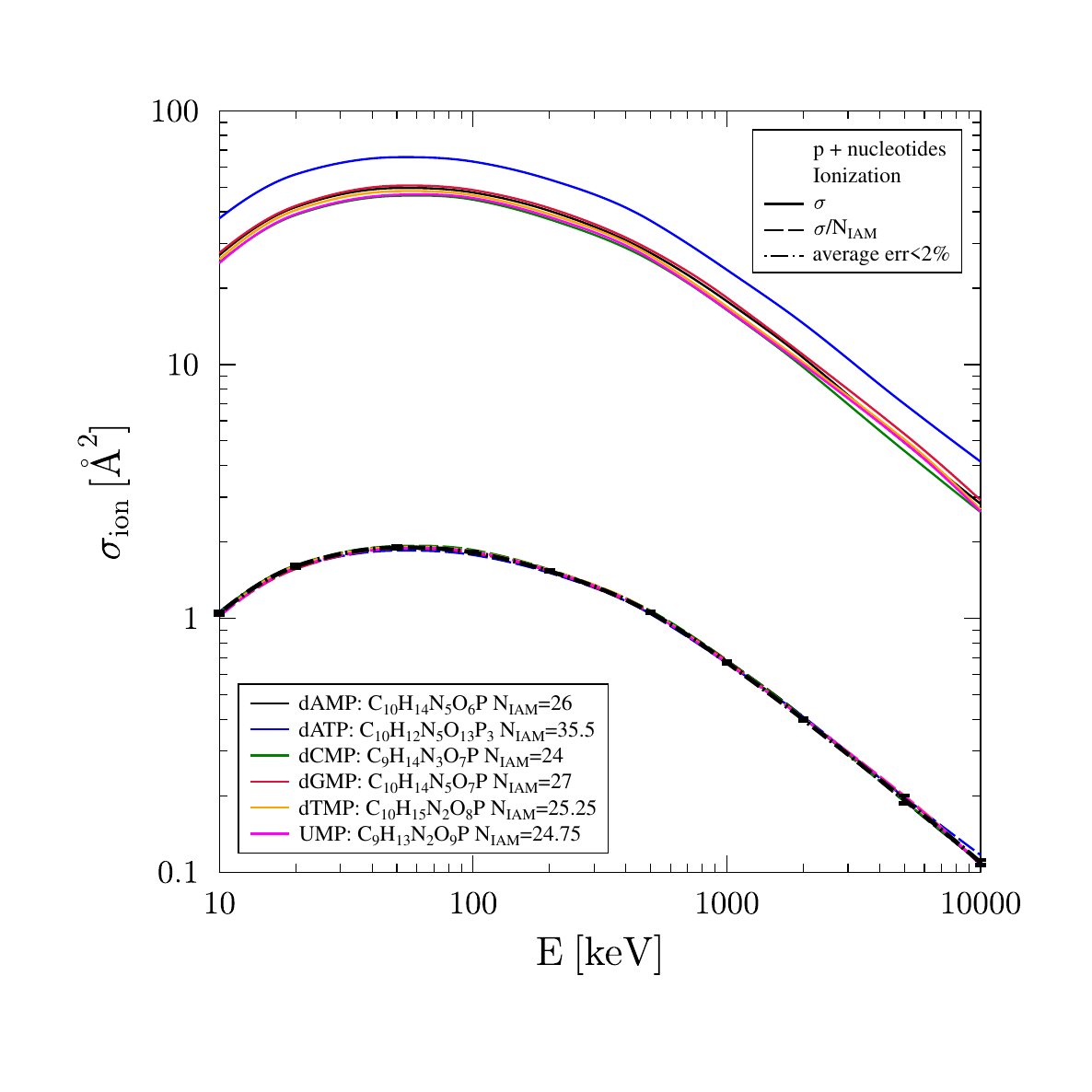}
}
\end{array}$
\vskip -0.5 truecm
\caption{Net ionization of amino acids (left panel) and nucleotides (right panel) by proton impact. 
Note the range of parameter values  $6.25 \le N_{\rm IAM} \le 12.25$ for the amino acids, and 
$24 \le N_{\rm IAM} \le 35.5$ for the nucleotides.}
\label{fig:Abb7ab}
\end{center}
\end{figure}

In the left panel of Fig.~\ref{fig:Abb7ab} we observe that among the presented amino acids, leucine ($\rm C_6H_{13}NO_2$, shown in red)
has the largest value of $N_{\rm IAM}$ and is at the top, while glycine ($\rm C_2H_5NO_2$, shown in green)
has the smallest $N_{\rm IAM}$ value and is at the bottom.
The ratio between the two curves is about two, i.e., it corresponds approximately to the ratio of values of $N_{\rm IAM}$.
The group of dashed curves for the reduced cross sections shows 
the excellent scaling behaviour of net ionization for amino acids.

Nucleotides are the building blocks of RNA and DNA.They are made up of one of the bases A, C, G, T or U, 
a monosaccharide and a phosphate group. 
The nucleotides of DNA differ from those of RNA by the absence of a hydroxyl group (OH) in the monosaccharide. 
To distinguish them, the nucleotides of DNA are called, for example, deoxyadenosine  monophosphate (dAMP). 
In addition, triphosphates are important for the energy supply of cells, especially ribosomes. 

In the right panel of Fig.~\ref{fig:Abb7ab} we observe that the results for dATP ($\rm C_{10}H_{12}N_5O_{13}P_{3}$, shown in blue)
stand out at the top among the net electron production cross sections for the group, which can be explained by the large
value of $N_{\rm IAM}=35.5$. The other nucleotides have similar values of $N_{\rm IAM} \approx 26$, and, thus,
their electron production cross sections as calculated with IAM-PCM are close to each other.
As a result the reduced cross sections form a narrow band as shown in the lower part of the right panel of Fig.~\ref{fig:Abb7ab}.

Electron production during proton impact is substantially larger for the nucleotides compared to the other biomolecular groups, 
as can be seen by comparing
the bands (in the top parts of Figs.~\ref{fig:Abb6ab}, and~\ref{fig:Abb7ab}). The reduced cross sections, on the other
hand, are very close to each other.
Whether the reduced cross sections for net ionization of these biomolecular groups can be understood using a simple
common model is discussed below.

We present our net electron production data for a large class of biological molecules to demonstrate that they can be
parametrized in a simple form. As mentioned in Sec.~\ref{sec:atomic-results}, at high impact
energies the atomic
net ionization cross sections effectively follow the form of the Bethe cross section
\begin{equation}
\sigma_{\rm Bethe}^{\rm ion}(E) = \frac{A \ln E + B}{E}  .
\label{eq:BB3}
\end{equation}
Here $A, B$ are parameters that can be obtained from fits to the calculated ionization cross sections,
and $E$ is the proton energy in keV.
The form of Eq.~(\ref{eq:BB3}) is obtained by analyzing the properties of individual ionized orbitals,
and proposals were made to apply the equation without calculating the generalized dipole oscillator 
strengths, e.g., recently in Ref.~\cite{PhysRevA.100.062702}. 
We  generalize the Bethe formula to lower energies where electron capture competes with
ionization. This can be accomplished by an effective projectile
charge parameter $Q_{\rm P}^{\rm eff}$ that depends on the collision energy~\cite{hjl19}:
\begin{equation}
\sigma^{\rm ion}_{\rm mod}(E)= [Q_{\rm P}^{\rm eff}(E)]^2 \sigma_{\rm Bethe}^{\rm ion}(E)   .
\label{eq:BB4}
\end{equation}
The effective charge is parametrized as
\begin{equation}
Q_{\rm P}^{\rm eff}(E) = 1 - \alpha \exp\{-(\beta E^{0.4})\}  ,
\label{eq:BB5}
\end{equation}
where the parameters $\alpha$ and $\beta$ are fitted for each of the previously discussed groups, and are given 
in Table~\ref{tab:BB1}. 
The fact that $\alpha > 1$ implies
that the model can only apply above some minimum proton energy $E$, but this is not a problem for $E>10 \rm \, keV$.

\begin{table}[ht]
{\begin{tabular}{@{}ccc@{}} \toprule 
{molecular group} &$\alpha$ & $\beta$ [$\rm keV^{-0.4}$] \\ \colrule
amino acids & 1.3450 &  0.19025 \\
pyrimidines & 1.3217 &  0.18316 \\
purines & 1.2933 &  0.17470 \\
nucleotides & 1.2563 &  0.16625 \\ \botrule
\end{tabular}}
\caption{Values of the parameters for the effective charge model Eq.~(\ref{eq:BB5}).}
\label{tab:BB1}
\end{table}

The total cross section for net ionization of a molecule from any given group can now be expressed as
\begin{equation} 
\sigma_{\rm mod}^{\rm ion}(E) =N_{\rm IAM} \left[ 1 - \alpha \exp\{-(\beta E^{0.4})\} \right]^2 \frac{100 \ln E + 120}{E} \, ,
\label{eq:BB6}
\end{equation}
where $N_{\rm IAM}$ is given in Eq.~(\ref{eq:BB1}), the proton collision energy $E$ is given in keV,
and the values of $A$ and $B$ in Eq.~(\ref{eq:BB3}) were taken from the observed
behaviour of the TC-BGM calculations for the atoms H, C, N, O, F.
The scaling of molecular cross sections takes a step further ideas that were applied in Ref.~\cite{hjl19}
at the level of atomic net ionization cross sections.
 
We can expect the ordering of the net ionization cross sections near the maximum to follow the pattern of the $\alpha$
and $\beta$ parameters, which vary in a systematic way.
The nucleotides in contrast with the other groups contain a phosphate, and have very large values of $N_{\rm IAM}$.
The function of the effective charge parameter $Q_{\rm P}^{\rm eff}(E)$ is to lower
the Bethe cross section gradually as the collision energy is below 1\,000 keV, and then dramatically
below 200 keV when capture begins to dominate. The difference in this effect between the molecular groups
is subtle but systematic, resulting in a spread of the net ionization cross sections at the maximum (between 40 and 100 keV)
at the level of ten percent.
 
\begin{figure}
\begin{center}$
\begin{array}{cc}
\resizebox{0.5\textwidth}{!}{\includegraphics{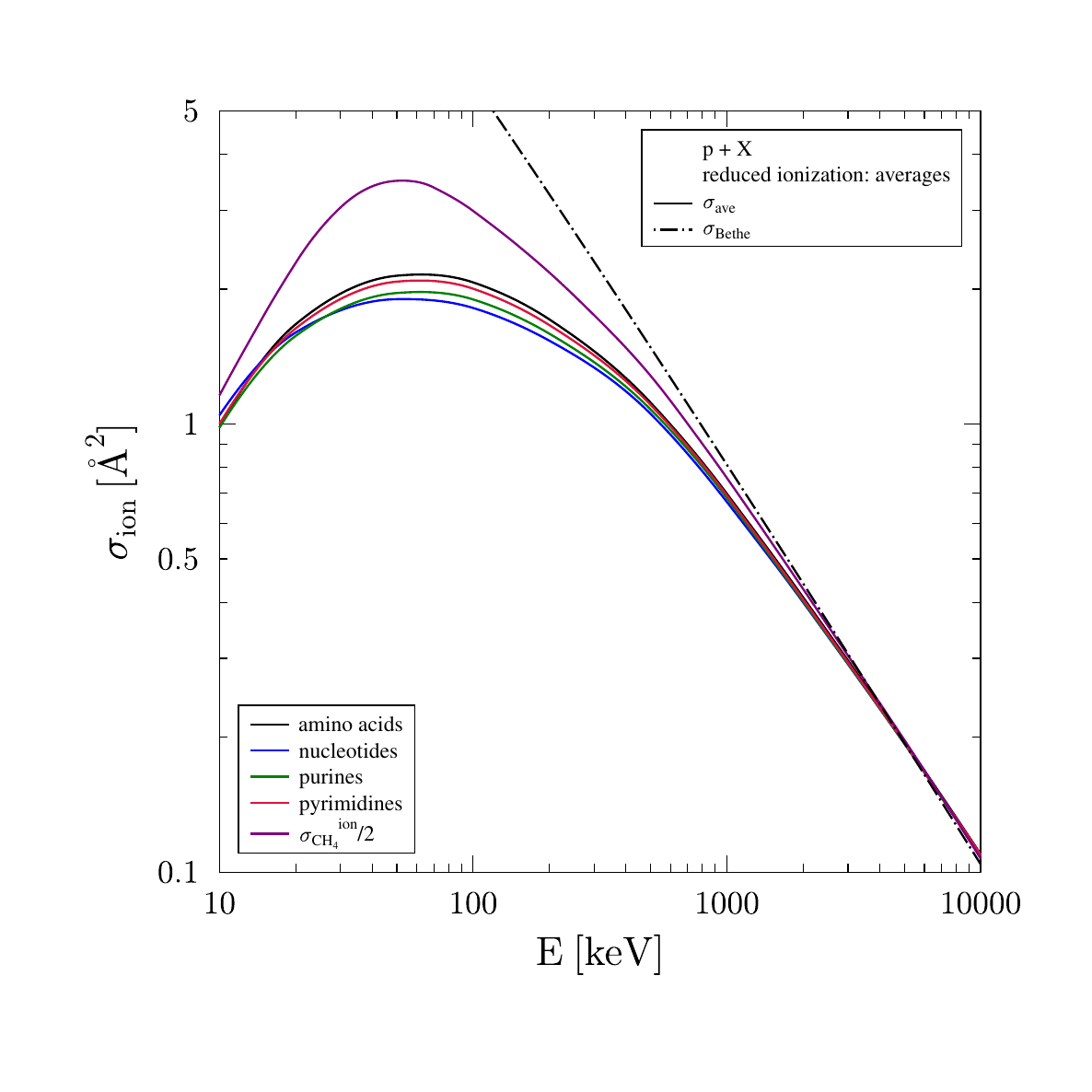}}&  \hskip - 0.6 truecm
\resizebox{0.5\textwidth}{!}{\includegraphics{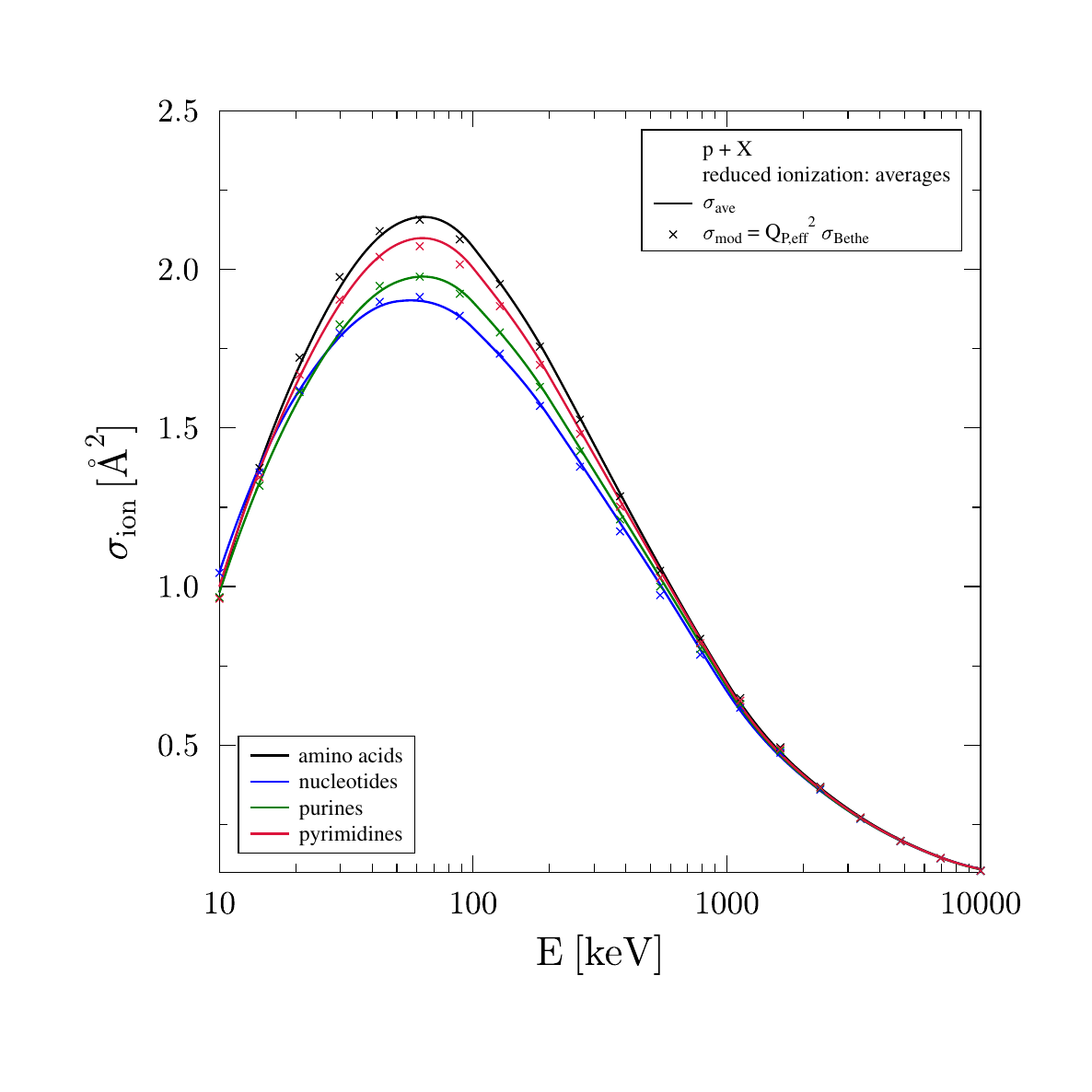}
}
\end{array}$
\vskip -0.5 truecm
\caption{Reduced net ionization cross section averages as shown in previous figures for the amino acids,
and nucleotides (Fig.~\ref{fig:Abb7ab}), purines and pyrimidines (Fig.~\ref{fig:Abb6ab}), and methane (Fig.~\ref{fig:Abb4ac}) are shown on a double-logarithmic scale in the left panel,
and (sans methane) on a log-linear scale in the right panel, which also shows how the curves pass through the data points
obtained from Eq.~(\ref{eq:BB6}).
}
\label{fig:Abb8ab}
\end{center}
\end{figure}

In Fig.~\ref{fig:Abb8ab} we demonstrate how well the scaling behaviour works over a large energy range.
The reduced cross section approach leads to a remarkably economic representation of net ionization
over the entire energy range and can be considered a success of the IAM-PCM. Of course, this is a theoretical
physics result, and we are awaiting eagerly confirmation by experiments. The challenge is also out there
for our theoretical colleagues, particularly those who perform truly molecular calculations.

Also shown in Fig.~\ref{fig:Abb8ab} is the reduced cross section for methane ($N_{\rm IAM}=2$).
It does not follow the scaling behaviour of the biomolecules, since the net electron emission peaks 
at a lower energy and at almost twice the value.

\section{Conclusions and Outlook}
\label{sec:conclusions}

The IAM-PCM has proven to be a capable and efficient approach to the calculation of capture and ionization 
cross sections in ion collisions with complex Coulomb systems such as molecules and clusters.
Originally introduced as a geometrical model, we
have argued in this chapter that it can be conceptualized in a different way by associating 
the pixels that make up the atomic cross section areas with scattering events. 
Ignoring multiple scattering events along a given projectile path 
is then equivalent to identifying
the molecular net cross sections as the combined area of overlapping atomic cross sections, i.e., the interpretation
we relied on in our previous work.

The new interpretation provides a more compelling rationale for the determination of impact-parameter-dependent
probabilities, as described in Sec.~\ref{sec:iam-pcm}, which form the basis for the calculation of charge-state correlated cross sections (Sec.~\ref{sec:iem}). 
Our results for proton--ten-electron collision systems summarized in Sec.~\ref{sec:10el} compare favourably with some 
but not all of the available experimental data. 
One conclusion from this is that there is a 
strong case for future complete coincident measurements of all contributing events for methane and
ammonia targets to provide a stringent test of the predicted universality of the cross sections ratios $\sigma_q/\sigma_1$.

We have carried out calculations for proton collisions with a large number of complex 
biomolecules. Comparisons with experimental data can be made for net cross sections and a few 
target systems such as uracil and adenine only.
As demonstrated in Sec.~\ref{sec:bio}, the IAM-PCM net electron production data exhibit a scaling behaviour that can be captured by a simple
parametrization with remarkable accuracy. We hope that this prediction will spark further experimental and theoretical
activity in the field, most importantly to advance the understanding of these complex collision systems, but also to help us
identify the limitations of the IAM-PCM.

This chapter has focused on proton-impact collisions, and we continue to study these systems. Our on-going work is also 
concerned with multply-charged projectile ions, and in some cases we have found that the IAM-PCM appears to
overcorrect additivity rule predictions of net cross sections. This raises the question whether the
single-scattering condition that is used as a model assumption needs to be relaxed.
We are currently looking into an extension of the PCM to allow for multiple scattering along a given projectile trajectory
using the concept of mean-free path as a criterion. 
Results will be reported in a future publication.

\begin{acknowledgments}
We would like to thank the Center for Scientific Computing, University of Frankfurt for making their High
Performance Computing facilities available. Financial support from the Natural Sciences and Engineering 
Research Council of Canada (Grants No. RGPIN-2023-05072 and No. RGPIN-2025-06277) is gratefully acknowledged.
\end{acknowledgments}

\bibliography{Chapter_hjmhtk}

\end{document}